\documentclass[%
aip,
amsmath,amssymb,
reprint,nofootinbib%
]{revtex4-2}
\usepackage{graphicx,array,booktabs,rotating}
\usepackage{dcolumn}
\usepackage{bm}
\usepackage{mathptmx}
\usepackage{multirow}
\usepackage{afterpage,float,color,xcolor}
\usepackage{hyperref}
\usepackage{etoolbox} 
\usepackage{tabularx,tabulary}
\usepackage[capitalize]{cleveref}
\usepackage{times}
\hypersetup{
	colorlinks,
	linkcolor={blue!100!black},
	citecolor={blue!100!black},
	urlcolor={blue!100!black}
}
\newcommand{\PRLsep}{\noindent\makebox[\linewidth]{\resizebox{0.3333\linewidth}{1pt}{$\bullet$}}\bigskip}
\makeatletter
\newcommand\footnoteref[1]{\protected@xdef\@thefnmark{\ref{#1}}\@footnotemark}
\makeatother
\renewcommand{\figurename}{FIGURE}
\newcommand{\sk}[1] {\textcolor{black}{#1}}
\newcommand{\etal}{\textit{et al.}}
\usepackage[page]{totalcount}
\usepackage{etoolbox,fancyhdr,xcolor}%
\usepackage[displaymath,pagewise]{lineno}

\begin{document}
\preprint{AIP/123-QED}
\title[\textbf{Journal to be decided} (2021) $|$  Revised Manuscript]{Shock and shear layer interaction in a confined supersonic cavity flow}

\author{S. K. Karthick}%
 \email{skkarthick@ymail.com}
\affiliation{ 
Faculty of Aerospace Engineering,\\ Technion-Israel Institute of Technology, Haifa-3200003, Israel
}%

\date{\today}

\begin{abstract}
The impinging shock of varying strengths on the free shear layer in a confined supersonic cavity flow is studied numerically using the detached-eddy simulation. The resulting spatiotemporal variations are analyzed between the different cases using unsteady statistics, $x-t$ diagrams, spectral analysis, and modal decomposition. \sk{A cavity of length to depth ratio $[L/D]=2$ at a freestream Mach number of $M_\infty = 1.71$ is considered to be in a confined passage.} Impinging shock strength is controlled by changing the ramp angle ($\theta$) on the top-wall. \sk{The static pressure ratio across the impinging shock ($p_2/p_1$) is used to quantify the impinging shock strength. Five different impinging shock strengths are studied by changing the pressure ratio: $1.0,1.2,1.5,1.7$ and $2.0$. As the pressure ratio increases from 1.0 to 2.0, the cavity wall experiences a maximum pressure of 25\% due to shock loading.} At [$p_2/p_1]=1.5$, fundamental fluidic mode or Rossiter's frequency corresponding to $n=1$ mode vanishes whereas frequencies correspond to higher modes ($n=2$ and $4$) resonate. Wavefronts interaction from the longitudinal reflections inside the cavity with the transverse disturbances from the shock-shear layer interactions is identified to drive the strong resonant behavior. Due to Mach-reflections inside the confined passage at $[p_2/p_1]=2.0$, shock-cavity resonance is lost. Based on the present findings, an idea to use a shock-laden confined cavity flow in an enclosed supersonic wall-jet configuration as passive flow control or a fluidic device is also demonstrated.
\end{abstract}

\maketitle

\section{Introduction}\label{sec:intro}
\sk{In a confined supersonic stream, shock-shear layer interactions are common and inevitable. They are encountered in variety of gas-dynamic scenarios including flow through ducts\cite{Li2017,Chakravarthy2018,Emmert2009,Afzal2020}, over-expanded nozzles\cite{Johnson2010,Olson2013,Chaudhary2020}, isolator of scram-jet engine\cite{Li2018,Devaraj2020,Sekar2020,RSaravanan2020}, enclosed open jet wind-tunnels \cite{Narayana2020,Sahoo2005,LePage2020}, and ejectors\cite{Karthick2016,Karthick2017,ArunKumar2016,Gupta2019}, where the influence of separated shear/boundary layer on the shock-laden flow is substantial.} One among such a flow field is confined supersonic cavity flow\cite{Pandian2018,Panigrahi2019,Kumar2018260,Li2013a}. Mixed-compression inlet for shock-wave boundary layer control\cite{Gefroh2003,Szulc2020}, resonator type supersonic nozzles\cite{Baskaran2019,Baskaran2021}, gas-dynamic lasers\cite{Schall1986,Shen1979}, and scram-jet combustors\cite{Tian2021,Oamjee2020} often employ cavities in supersonic flow which are placed in closed proximity to the surrounding walls. Aerodynamic testing facilities like the wind-tunnel facilities aim to study the multi-purpose bay-door interaction with freestream, as seen in supersonic aircraft, by modeling the system as a cavity on the test-section wall \cite{Thangamani2019,Zhang1988365,Lad2018,Xiansheng2020}. Flow past such cavities often suffer from wall effects. These interference change the fluid dynamics drastically from the actual encounters, for instance, in actual flights, where the effects of walls are completely or partially absent.

Supersonic cavity flows exhibit emission of distinct tones in multiple frequencies, primarily due to the feedback mechanism that drives the vortex shedding through the cavity shear layer\cite{Rockwell1978,Heller1971,Kegerise2004,Li2013b}. The traveling waves inside the cavity and sound production due to vortex bursting in the cavity's trailing edge contribute to the feedback. These events happen in an organized way which enables the shear layer to oscillate in a self-sustained manner. Pressure oscillations due to the presence of a cavity dictate the dominant flow characteristics in a duct. The dominant temporal modes' behavior governs the dynamics in holding a flame in a combustor, unsteadiness in the isolator section of the supersonic inlet, and resonator nozzle control efficiency.

Cavity flows have been studied both experimentally\cite{Cattafesta2008,Handa20113383,Unalmis2001} and computationally\cite{Barnes2015,Rizzetta2008,Lawson2011} in a variety of flow regimes from different points of view ranging from aeroacoustics\cite{Malhotra2016411,Gautam2017211,Maurya2015559,Vikramaditya20121389} to heat transfer\cite{CHARWAT1961,Emery1969,Batov1998,Vishnu2019}. However, the majority of the outcomes from the computational analysis related to confined cavity flows assume clean flow in the freestream\cite{Li2008,Rizzetta1988,Barakos2009,Jian2021} i.e., flow without any shocks or compression waves (Mach waves). On the contrary, shocks in the confined flow are unavoidable due to various factors, including surface irregularities and shock-wave turbulent boundary layer interactions. Even, in reality, these inherent contributions from confinement are inexorable. In the scramjet combustor, shock-induced combustion\cite{Urzay2018,Kirik2014,Etheridge2017} is achieved through shock-on-jet/cavity configuration to ignite the air-fuel mixture. \sk{Generally, theoretical and computational cavity flow analysis only assumes the in-flow with no-disturbances.}

\begin{figure}
	\includegraphics[width=0.95\columnwidth]{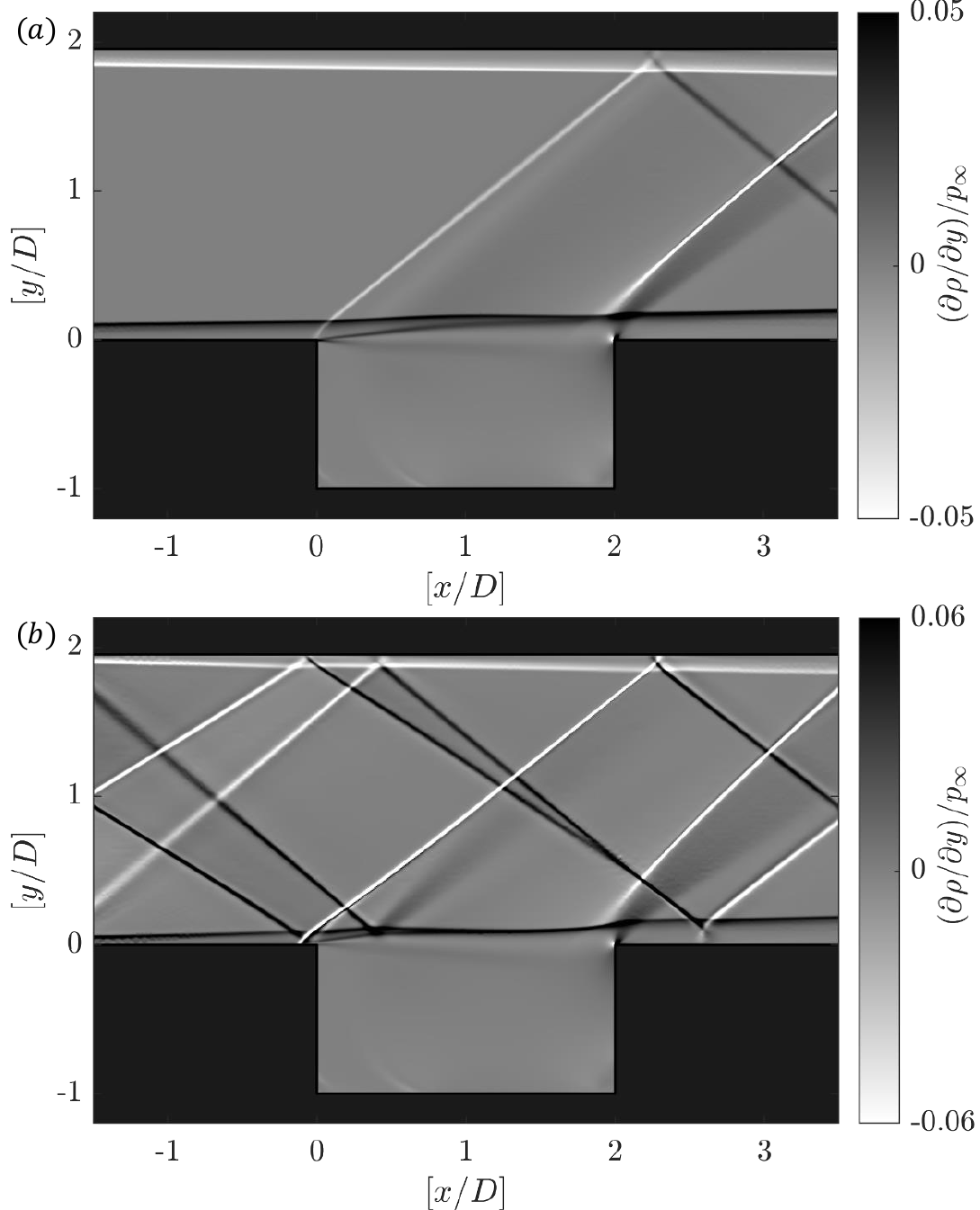}
	\caption{\label{fig:gen_diff} Non-dimensionalized numerical schlieren based on the density gradients along the $y$-direction, $\left(\partial \rho/\partial y\right)/p_\infty$, showing the differences between a confined supersonic cavity flow (a) without and (b) with upstream disturbances (produced through a pair of tiny protrusion ramps $(\sim0.04D)$ on either side of the wall at $[x/D]=-8$) using a RANS based simulation. Flow is from left to right. Flow parameters: $M_\infty=1.71$, $p_\infty=1.01 \times 10^5$ Pa, $T_\infty=189.3$ K.}
\end{figure}

A typical normalized numerical schlieren along the flow direction without and with freestream disturbances in the form of compression waves is shown in Figure \ref{fig:gen_diff} for a confined supersonic cavity flow using a RANS (Reynolds Averaged Navier Stokes) based simulation. In the clean flow (Figure \ref{fig:gen_diff}a), the dominant flow features around the cavity, including the separation shock, shear layer, and reattachment shock, are completely different than what is observed in the flow with disturbances (Figure \ref{fig:gen_diff}). The freestream disturbances or the presence of roller structures in the turbulent boundary layer create wavefronts. These wavefronts interact with the near-wall boundary layer in the confined passage and induce weak or strong shock-wave boundary-layer interactions (SWBLI) upstream of the cavity. The reflecting waves from the walls interact with the separated free shear layer of the cavity and alter the shear layer's physical morphology. The implications of such distortions around the cavity due to shock-shear layer interactions might either be weak or strong based on the impinging shock strength and the shear layer thickness.

For a long time, the observed flow dynamics' actual causality due to freestream disturbances was not a concern as the major outcomes were within the experimental deviations. Many experimental facilities do not report the disturbances or quantify their values, as their focus is to solve an applied engineering problem. However, a systematic analysis on the influence of quantified freestream disturbances in the form of the impinging shock wave on the cavity shear layer is not available in the open literature to the author's best of knowledge. Hence in the present paper, the author explores the spatiotemporal changes in a confined supersonic cavity flow, particularly due to shock-shear layer interactions. Following are the distinct objectives of the present paper:
\begin{enumerate}
    \item {To numerically simulate the unsteady flow past confined supersonic cavity with different strengths of shock-shear layer interactions.}
    \item{To compare the unsteady statistics on the cavity-wall between the different cases to ascertain the influence of shock impingement on the cavity dynamics}.
    \item{To perform a spectral analysis along the cavity wall and also in the freestream to identify any coupling or resonant behavior between the cavity oscillations and the shock motions in confinement}.
    \item{To identify the dominant spatiotemporal mode changes between the different shock-impingement cases using modal analysis tools like DMD (dynamic mode decomposition)}.
\end{enumerate} The outcome of the present paper will dictate the permissible limits of the impinging shock strength on the cavity shear layer without changing the cavity dynamics. The findings will also assist to carefully interpret the unsteady measurements from the future shock-laden cavity flow experiments. 

The paper is organised as follows. In $\S$\ref{sec:problem}, the problem statement is briefed with geometrical configurations and flow conditions. Numerical methodology is discussed in $\S$\ref{sec:numerical}, where the details about the computational domain, flow solver, grid/time independence study and steady/unsteady validations are provided. In $\S$\ref{sec:res_disc}, results and discussions from the steady and unsteady statistics, cavity shock-footprint analysis, shock oscillation spectra in the duct, and dominant temporal modes are given. \sk{In $\S$\ref{sec:application}, the possibility of using the shock-laden cavity as a passive control device is explored. Present computations limitations, and future scope for further research are discussed in $\S$\ref{sec:scope}.} In the last section ($\S$\ref{sec:conc}), major outcomes of the present study are summarized.

\section{Problem Statement}\label{sec:problem}
\begin{figure}
	\includegraphics[width=\columnwidth]{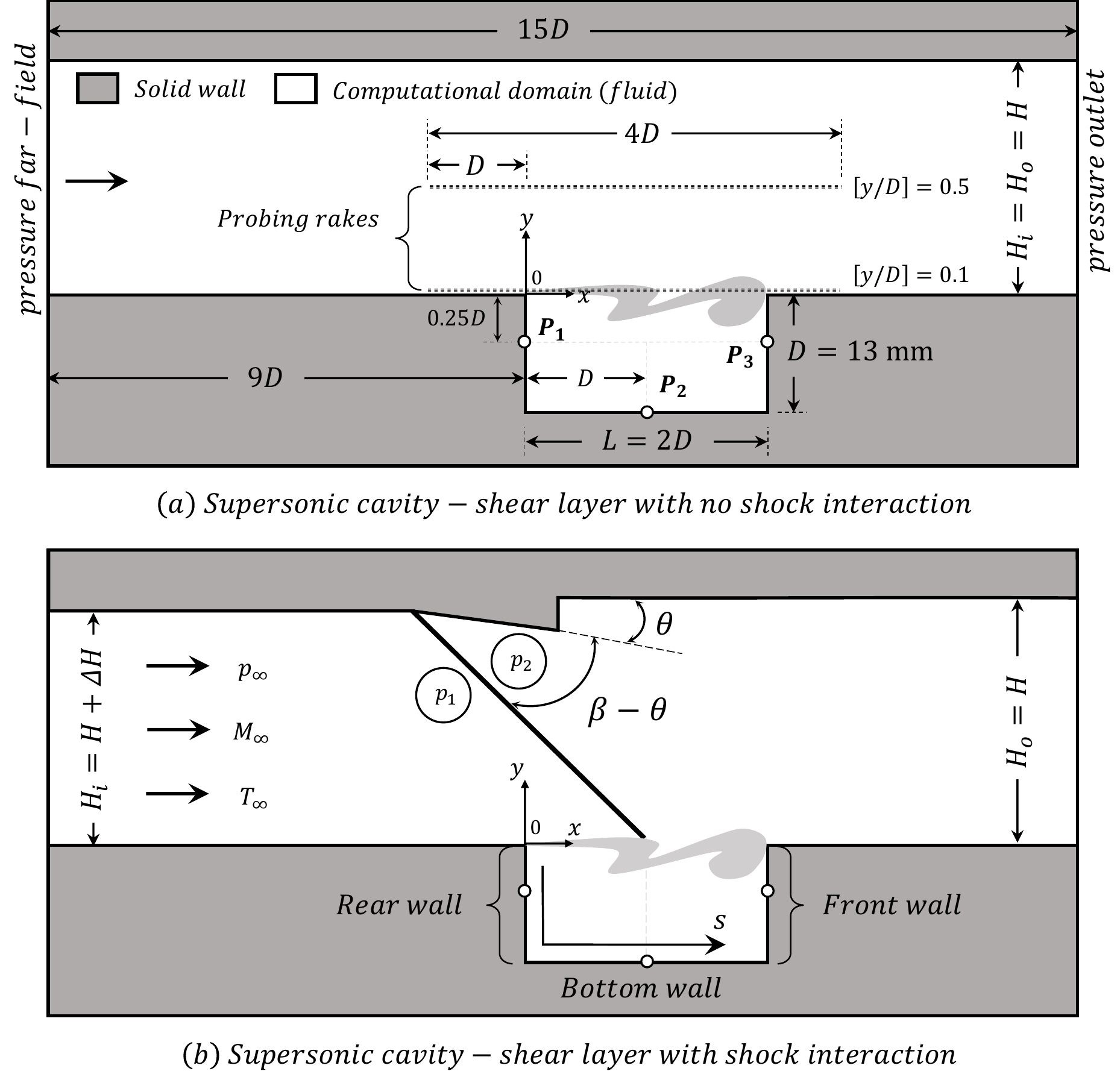}
	\caption{\label{fig:prob_statement} \sk{Schematic describing the problem statement, computational domain extent, numerical boundary conditions, freestream direction, and probing stations for flow past a confined supersonic cavity: (a) without and (b) with shock-shear layer interactions of varying strength. Flow is from left to right. The schematic is not drawn to the scale.}}
\end{figure}

In the present numerical study, a problem statement has been formulated to compare the dynamics inside a confined cavity with no shock interactions on the cavity shear layer with the cases of different impinging shock strengths (see Figure \ref{fig:prob_statement}). The shock is generated using a simple ramp placed on the top wall of the confinement. Ramp angles ($\theta$) are changed to achieve different shock strengths. Due to changes in $\theta$, the shock angle ($\beta$) also changes (see Table \ref{tab:case_list}). \sk{The ramp on the top-wall is moved accordingly in the streamwise direction. Thus, the shock interaction point on the cavity shear layer almost remains the same.} The ramp shock's ideal interaction point is set to be at $[x/D] \sim 1$ along the cavity shear layer. Shock strength is quantified by the pressure jump across the shock ($p_2/p_1$). Including the no-shock condition, five different cases are formulated with varying shock strengths: $[p_2/p_1]=$1.0 (no shock-interaction), 1.2, 1.5, 1.7, and 2.0 to represent a broad range of cases from mild to strong shock-shear layer interaction as shown in Table \ref{tab:case_list}. The total pressure loss ($p_{02}/p_{01}$) incurred due to the ramp shock reduces the mass flow rate in the duct. Such losses are corrected by changing the inlet height slightly ($H_i=H+\Delta H$, where $H$ is the duct height in the no-shock interaction case) and by keeping the outlet height at a constant value ($H_o=H$). Thus, the net mass flux across the confined passage remains the same between the cases. 

\subsection{Geometrical configuration}\label{sec:geometry}
Geometrical configuration utilized in the experimental work of Kumar and Vaidyanathan \citep{Kumar2018260} is considered here. The cavity is of the open type with a length ($L$) to depth ($D$) ratio of $[L/D]=2$. The confinement's upstream distance is extended to $[x/D]=9$ to attain a developed boundary layer of considerable thickness ($\delta$). A regime spanning $[x/D]=4$ is considered for the analysis downstream of the cavity. The flow is from left to right, with the origin located on the cavity leading edge. Any point on the cavity surface could be accessed using the surface length coordinate system given by $s$. The details of the geometrical configurations are given in Figure \ref{fig:prob_statement}. Measurements are taken at different locations. \sk{Unsteady statistics are collected at three probing locations along the cavity wall ($[s/D]=0.25,2,3.75$), which are later utilized to understand the flow physics and also to compare with the findings of Panigrahi \etal\citep{Panigrahi2019}.} Two streamwise rakes extending between $-1 \leq [x/D] \leq 3$ are placed at differed transverse locations ($[y/D]=$0.1, and 0.5) in the freestream to understand the influence of cavity dynamics on the shock oscillations. 

\begin{table}
	\caption{\label{tab:case_list}\sk{Different types of cases realized in the numerical study by varying the ramp-angle ($\theta$), shock-angle ($\beta$), ratio of the inlet duct-height and cavity depth ($H_i/D$) and the impinging shock-strength ($p_2/p_1$, see Figure \ref{fig:prob_statement} for the definition) in a confined supersonic cavity flow.}}
	\begin{ruledtabular}
		\begin{tabular}{ccccc}
			Configurations & $\theta$ $(^\circ)$ & $\beta$ $(^\circ)$ & $[H_i/D]$ & $[p_2/p_1]$ \\ 
			\midrule 
			C-0\footnote{Confined supersonic flow past a rectangular cavity without any ramps or upstream disturbances to produce shock-shear layer interaction.} & 0 & 0 & 1.954 & 1 \\
			C-1  & 3.6 & 39.3 & 1.958 & 1.2 \\
			C-2  & 8.2 & 44.5 & 1.969 & 1.5 \\
			C-3  & 10.65 & 47.7 & 1.984 & 1.7 \\
			C-4  & 13.78 & 52.87 & 2.021 & 2.0 \\
		\end{tabular}
	\end{ruledtabular}
\end{table}

\subsection{Flow conditions}\label{sec:flow cond}
Present analyses are carried out at a moderate supersonic freestream Mach number of $M_\infty=1.71$. The freestream's total pressure and temperature are taken as $p_0=5$ bar and $T_0=300$ K, respectively. The boundary layer ($\delta/D$) and momentum thickness ($\theta/D$) at the entrance of the cavity ($[x/D]=0$) is 0.15 and 0.015. All other vital flow parameters are tabulated in Table \ref{tab:flow_cond}.

\begin{table}
	\caption{\label{tab:flow_cond} Tabulation of freestream flow conditions achieved ahead of the confined supersonic cavity with/without shock-shear layer interactions.}
	\begin{ruledtabular}
		\begin{tabular}{lc}
			Parameters & Values \\ 
			\midrule 
			Cavity length ($L$, $mm$) & 26 \\
			Cavity depth ($D$, $mm$) & 13 \\
			Boundary layer thickness ($\delta_{x=0}$, $mm$) & 1.9 \\
			Momentum thickness ($\theta_{x=0}$, $mm$) & 0.2 \\
			Freestream velocity ($u_\infty$, $m/s$) & 471.6 \\
			Freestream static pressure ($p_\infty\times 10^5$, $Pa$) & 1.01 \\
			Freestream static temperature ($T_\infty$, $K$) & 189.3 \\
			Freestream static density ($\rho_\infty$, $kg/m^3$) & 1.86 \\
			Freestream kinematic viscosity ($\nu_\infty\times 10^{-6}$, $m^2/s$) & 6.84 \\
			Freestream Mach number ($M_\infty$) & 1.71 \\
			Reynolds number based on $\theta$ ($Re_\theta \times 10^4$) & 1.33 \\
		\end{tabular}
	\end{ruledtabular}
\end{table}

\begin{figure*}
	\includegraphics[width=\textwidth]{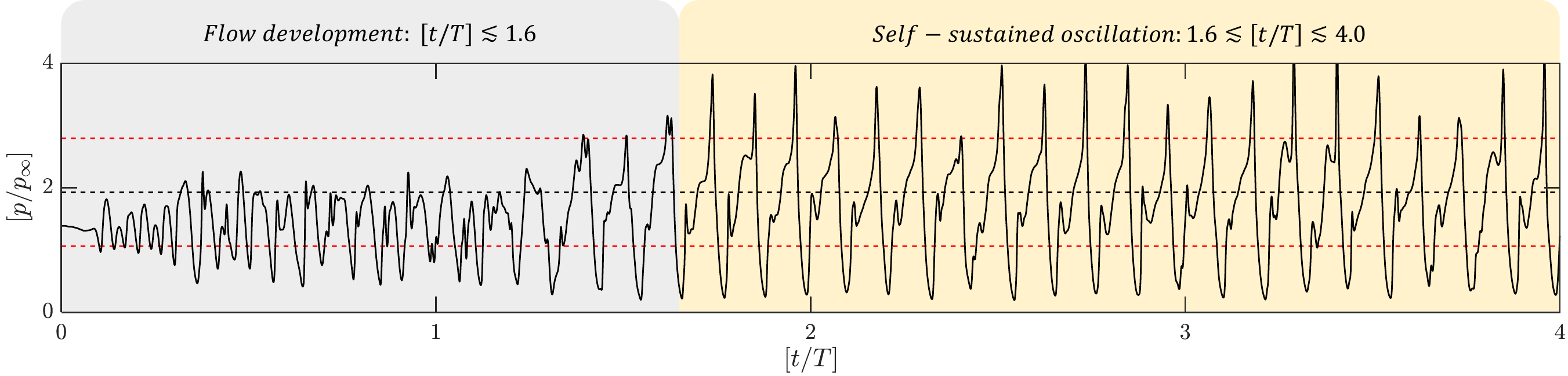}
	\caption{\label{fig:unsteady_signal} A typical unsteady wall-static pressure profile ($p/p_\infty$) observed at $P_3$ probing location at $[s/D]=3.75$ in the $[p_2/p_1]=1.5$ configuration. The flow development time required during a numerical run and the extent of time where the self-sustained oscillation regime is observed due to shock-shear layer interaction in a confined supersonic cavity flow is shown. Black-dotted line shows the $rms$ and red-dotted line show the standard deviation about the $rms$ during self-sustained oscillation (Note: The global time stamp is locally reset to zero once the self-sustained oscillation begins for plotting convenience in the upcoming figures. Hence, the local time stamps given in them should not be compared with the global time stamp mentioned in the present figure unless otherwise specified).}
\end{figure*}

\section{Numerical Methodology}\label{sec:numerical}

\sk{A commercial computational fluid dynamics solver package from Ansys-Fluent\textsuperscript{\tiny\textregistered} is used to investigate the transient flow past the two-dimensional confined supersonic cavity. The fluid turbulence is modeled using a hybrid RANS/LES (Large Eddy Simulation) scheme called Detached Eddy Simulation\cite{Ansys} (DES). Following are the major reasons\cite{Menter2021} for the selection of DES schemes in contrast to the traditional URANS (unsteady RANS) or high-fidelity LES approaches: I. URANS suffers to predict flows dominated by global instabilities faithfully, II. LES pose stringent requirements on near-wall grid preparation and is computationally expensive. On the other hand, DES resolves the unsteady turbulent features in massively separated flows like the cavity flow at low computational costs and offers flexibility in grid generation\cite{Andreini2016}. While using DES, the turbulent structures away from the wall like the separated regions are modeled using the LES-like subgrid models\cite{Rizzetta2008}. Wall-bounded turbulence is modeled using a traditional RANS\cite{Anavaradham2004} schemes. The cavity shear-layer features like the transient vorticity including the origin, and formation of a scroll was modeled using DES schemes, previously\cite{Liu2015,Lawson2011}.} Hence, in the present studies, a similar numerical approach is adopted to access the spatiotemporal dynamics across the different types of shock-shear layer interactions in a two-dimensional confined supersonic cavity flow.

\subsection{Boundary conditions and solver settings}\label{sec:solver}

\sk{A two-dimensional\footnote{\sk{A three-dimensional RANS simulation is performed for the $[p_2/p_1]=1.0$ case (no shock-interaction) and the results are compared with the two-dimensional counterpart. The centre-line ($z/D=0$) cavity wall-static pressure from the three-dimensional grid ($5 \times 10^6$) only deviates to a maximum value of 5\% in comparison to the solutions from the two-dimensional grid ($1 \times 10^5$). Hence, the author did not seek high-fidelity computations (like 3D-DES or 3D-LES) separately using a three-dimensional grid. However, the author would like to emphasize that only the dominant longitudinal flow modes are resolved in the considered two-dimensional grid which might be sufficient to describe the causality of shock-shear layer interactions as described in the upcoming sections. The confounding effects of the cavity's three-dimensionality may also play an important role but they are not considered in the present study as it is beyond the scope of current investigation.}} planar computational domain is considered with spatial extents as mentioned in $\S$ \ref{sec:geometry}. Pressure far-field and pressure outlet boundary conditions are given at the domain entrance and exit, respectively (see Figure \ref{fig:prob_statement}a and Table \ref{tab:flow_cond}). The inlet turbulence is specified through turbulent intensity ($5\%$) and turbulent viscosity ratio (10).} Air at ideal-gas condition is used as the working fluid. Fluid viscosity is modeled using Sutherland's three-coefficient method. The near-wall turbulence is modeled through RANS $k-\omega$ SST\cite{Menter1994} (shear-stress transport) along with the DDES\cite{Allen2004} (delayed-DES) shielding function with default constants given in the solver. The surrounding solid-walls are kept at adiabatic conditions. 

\sk{A coupled pressure-based solver with a compressibility correction is preferred in contrast to the traditional density-based solver for high-speed flows. In recent days, the coupled pressure-based solver is proven to predict the supersonic\cite{Chaudhary2020} or hypersonic\cite{Sekar2020} flow field with a large region of low $Re$ flow in the domain reasonably well. Besides, it exhibits solution acceleration and stability. Between the two solvers, the maximum deviation in the computed mean wall-static pressure ($\overline{p}/p_\infty$) along the cavity wall ($s/D$) is identified to be merely $3.5\%$ for the $[p_2/p_1]=1.0$ case (no shock-interaction). Hence, the coupled pressure-based solver with compressibility correction is incorporated to swiftly simulate the steady-state solution. Firstly, the steady-state solution is achieved through hybrid initialization before switching to transient. Hybrid initialization is favored again for rapid solution convergence. It predicts the values in the fluid domain almost closer to the expected through internal iterations (typically in 10 iterations) by solving a simplified system of flow equations\cite{Ansys}. The gradients are spatially discretized using a Green-Gauss node-based method with second-order upwind schemes except for the momentum terms in the equations, which uses a bounded-central differencing. Second-order implicit scheme with iterative time-advancement scheme is used to achieve the transient formulation and to eliminate the splitting errors while segregating the solution process.} 

The present computations take $[t/T]\lesssim 1.6$ for the unsteady flow to develop in the computational domain ($T=1$ ms, reference time). For $[t/T] > 1.6$, a self-sustained oscillation in the confined cavity is generally seen for all the considered cases. Typically, computations are performed until 20 dominant oscillation cycles are captured. For example, in configuration $[p_2/p_1]=1.5$ (C-2 case in Table \ref{tab:case_list}), the flow development and self-sustained oscillation time zones are marked in Figure \ref{fig:unsteady_signal} while monitoring the pressure fluctuations at $P_3$ location in the cavity's front-wall. A separate time-step independence study is done in $\S$\ref{sec:time_Step_ind}. Based on the results, a global time-step size of $[t/T]=1 \times 10^{-3}$ is selected. For each time step, 50 iterations are carried out and a convergence of $10^{-5}$ is achieved in the local scaled-residuals of the continuity equation. 

\subsection{Grid independence and validation}\label{sec:grid ind}

\begin{figure}
	\includegraphics[width=0.9\columnwidth]{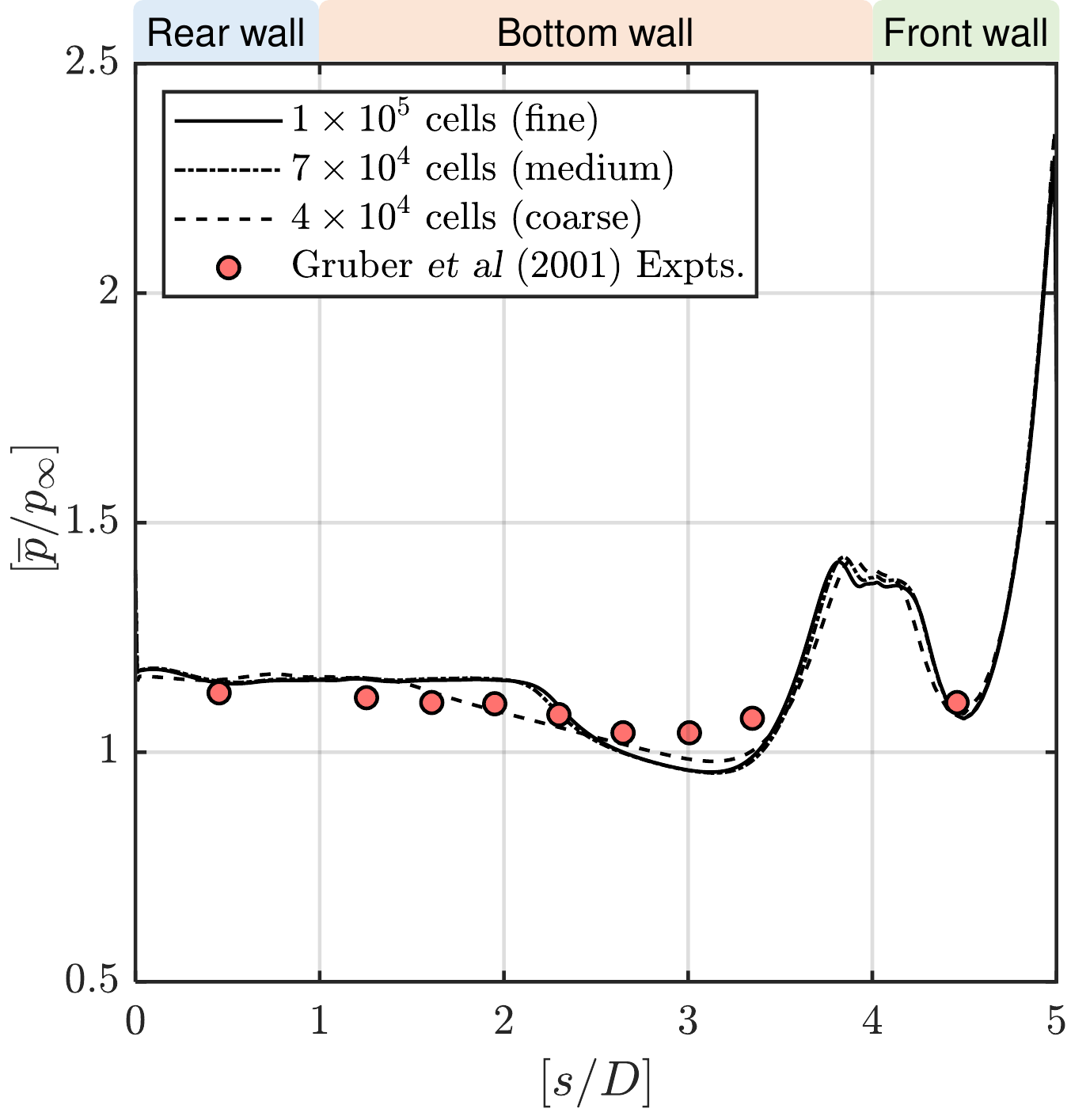}
	\caption{\label{fig:pressure_validation} Validation of present computation grid and flow solver in RANS mode with the experiments through the comparison wall static pressure ($p/p_\infty$) along the cavity wall surface ($s/D$) for the following freestream conditions: $[L/D]=3.0$, $M_\infty=3.0$, $p_0=690$ $kPa$, and $T_0=300$ K.}
\end{figure}

Grids are generated using a commercial software called Ansys-Gambit\textsuperscript{\tiny\textregistered}. A two-dimensional fluid domain is meshed using structured-quadrilateral cells. More than 96\% of the cells contain an equisize skewness parameter of 0.2. The turbulence wall parameter is maintained at $y^+<1$ everywhere to resolve the boundary layer. The streamwise and transverse cell-size progression is maintained not more than 1.2 across the fluid domain, especially closer to the cavity, to capture the shock motions faithfully. Three types of the grid of different grid densities are considered to study the solution independence to grid density changes: (a) coarse grid ($4\times 10^4$ cells), (b) medium grid ($7\times 10^4$ cells), and (c) fine grid ($1 \times 10^5$ cells). The total number of cells is varied based on the grid density in the vicinity of the cavity. Steady-state wall-static pressure ($\overline{p}/p_\infty$) is monitored along the cavity walls for all the different grids and plotted in Figure \ref{fig:pressure_validation} for an experimental case of Gruber \etal \cite{Gruber2001146} towards validation. \sk{The computed $[\overline{p}/p_\infty]$ values from the coarse grid matches almost well with the experiments. Between the medium and fine grid, the numerical solution does not deviate much indicating grid independence. However, only fine grid is considered for all the simulations as it is independent and sufficient to capture the shocks without further grid adaption which is computationally costly.} 

\sk{The deviations of numerical solution from the experiments can be argued in two fronts: a. dimensionality of the problem, and b. uncertainties in the experiments. The present simulation is only two-dimensional. If any lateral instabilities due to three dimensional effects arise (due to sidewalls, translational/rotational periodicity, or unbounded lateral openings), then the associated change in flow physics will not be captured in the present simulation and the results deviate from the experimental findings. Besides, the experimental values also have inherent uncertainties. Including the sensors (0.05\%, as reported by Gruber \etal\cite{Gruber2001146}), additional uncertain features like the pressure taps (of about 1 mm), the A/D (analogue to digital) sampler, and data acquisition system need to be considered while computing the overall uncertainty, which is not available in the considered experimental report. In similar compressible flow wall-static pressure measurements as seen in the literature\cite{Rao2014,Karthick2018}, the combination of factors mentioned above adds to a total uncertainty of about $\pm 5\%$. Based on the dimensionality constraints and the experimental uncertainties in the validation case, the fine grid is considered suitable for further numerical studies.}   

\subsection{Time-step independence and validation}\label{sec:time_Step_ind}

\begin{figure}
	\includegraphics[width=\columnwidth]{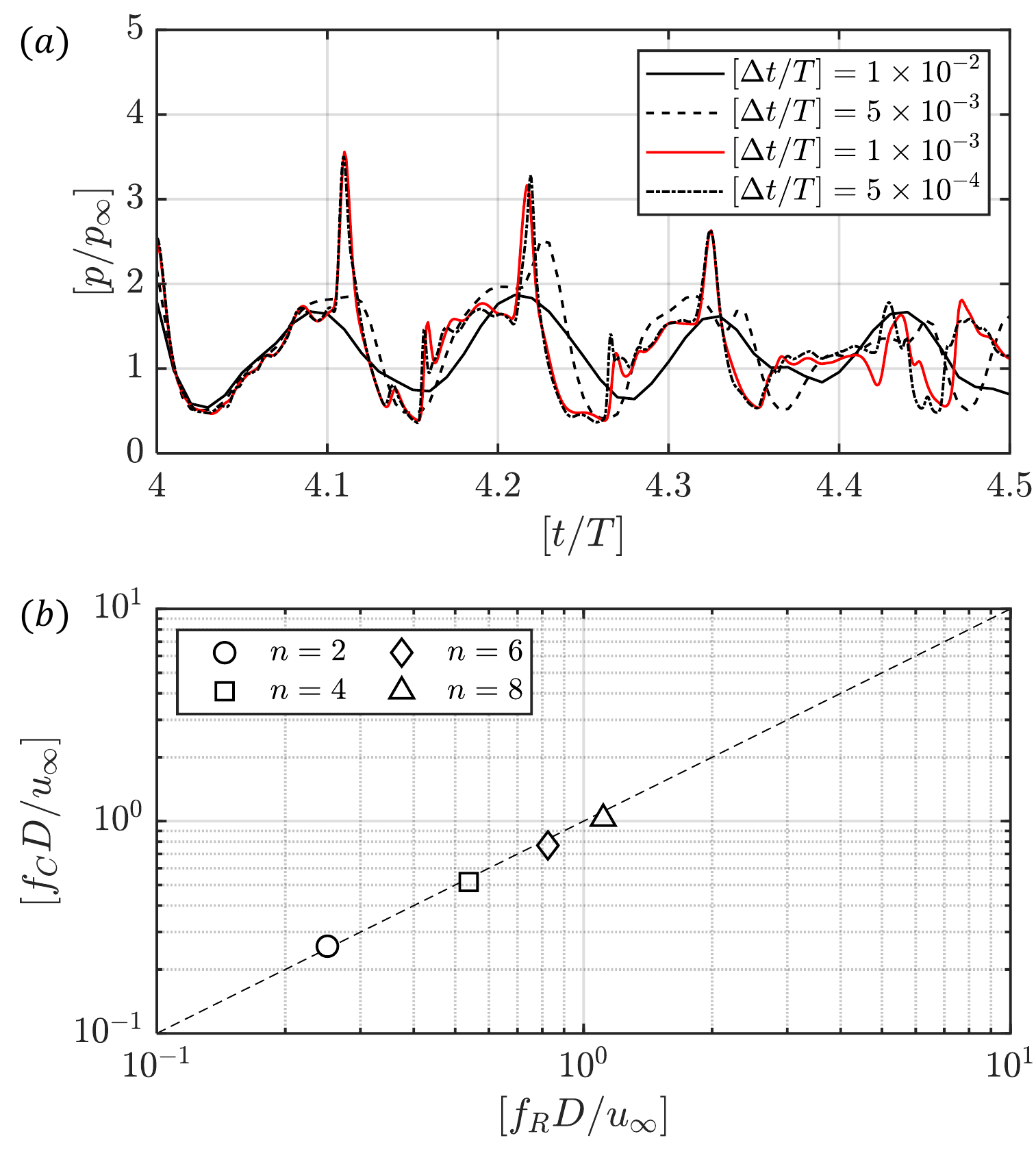}
	\caption{\label{fig:time_step_validation} (a) Time-step independence studies for the DES simulation carried out at different $[\Delta t/T]$ by monitoring the time variation of [$p/p_\infty$] at $P_3$ location while capturing at-least three dominant cycles using the fine grid ($1 \times 10^5$ cells) and keeping rest of the solver parameters the same. (b) Validating the computed dominant frequencies ($f_C$) from the moderately fine time-step solution of $[\Delta t/T]=1 \times 10^{-3}$ ($T$=1 ms) containing 20 dominant unsteady cycles with the calculated frequencies from the modified Rossiter equation ($f_R$, see Eq.\ref{eq:ross_eqn}) for particular modes ($n=2,4,6,8$) by assuming suitable values for the constants ($k=0.47,\alpha=0.25$) as suggested by Heller \etal\cite{Heller1971} and Kegerise \etal\cite{Kegerise2004}.}
\end{figure}

A time-step independence study is performed on a plain cavity flow to ensure that the well-defined Rossiter modes are captured at a sufficiently resolved temporal resolution or time-step. Four non-dimensionalized time-steps are considered ($\Delta t/T$): (a) $1 \times 10^{-2}$, (b) $5 \times 10^{-3}$, (c) $1 \times 10^{-3}$, and (d) $5 \times 10^{-4}$. Different time-steps are selected based on the physical limits to resolve the dominant frequency, utilized time-steps as mentioned in the literature, and the requirement of computation resources. The wall-static pressure at $P_3$ probing point (see Figure \ref{fig:prob_statement}) is monitored and at-least three dominant $[p/p_\infty]$ cycles are captured at each time-step for comparison in Figure \ref{fig:time_step_validation}a. The finest ($\Delta t/T=5 \times 10^{-4}$) and the moderately fine \sk{($\Delta t/T=1 \times 10^{-3}$)} time-step considered in the current exercise almost overlap one over the other, whereas the other two time-steps are identified to produce a completely different trend. The moderately fine time-step \sk{($\Delta t/T=1 \times 10^{-3}$)} is selected for the rest of the numerical exercise as it fairly captures the dynamics and reduces the computation load. 

\sk{The wall-static pressure signal $(p/p_\infty)$ containing 20 dominant unsteady cycle as seen in Figure \ref{fig:unsteady_signal} at $P_3$ location is subjected through the fast-Fourier transform (FFT) analysis.} The computed dominant frequencies are plotted against the respective frequencies as given by the modified Rossiter formula \cite{Heller1971} (see Eq.\ref{eq:ross_eqn}, where $n$-mode number, $\alpha$-phase delay between the acoustic wave and vortex generation, $k$-ratio of the vortex convection speed to the freestream). A Maximum of 2.5-7.5\% variation is seen between the computed and calculated dominant frequencies of certain modes ($n=2,4,6,8$) for the assumed values of $\alpha=0.25$ and $k=0.47$ as suggested by Heller \etal\cite{Heller1971} and Kegerise \etal\cite{Kegerise2004}. The relevant validation plot is shown in Figure \ref{fig:time_step_validation}b.

\begin{equation}
    \frac{f_R D}{u_\infty} = \frac{n-\alpha}{M_\infty\left(1+[(\gamma-1)/2]M_\infty^2\right)^{-1/2}+1/k} 
    \label{eq:ross_eqn}
\end{equation}

\section{Results and Discussions}\label{sec:res_disc}

\begin{figure*}
	\includegraphics[width=1\textwidth]{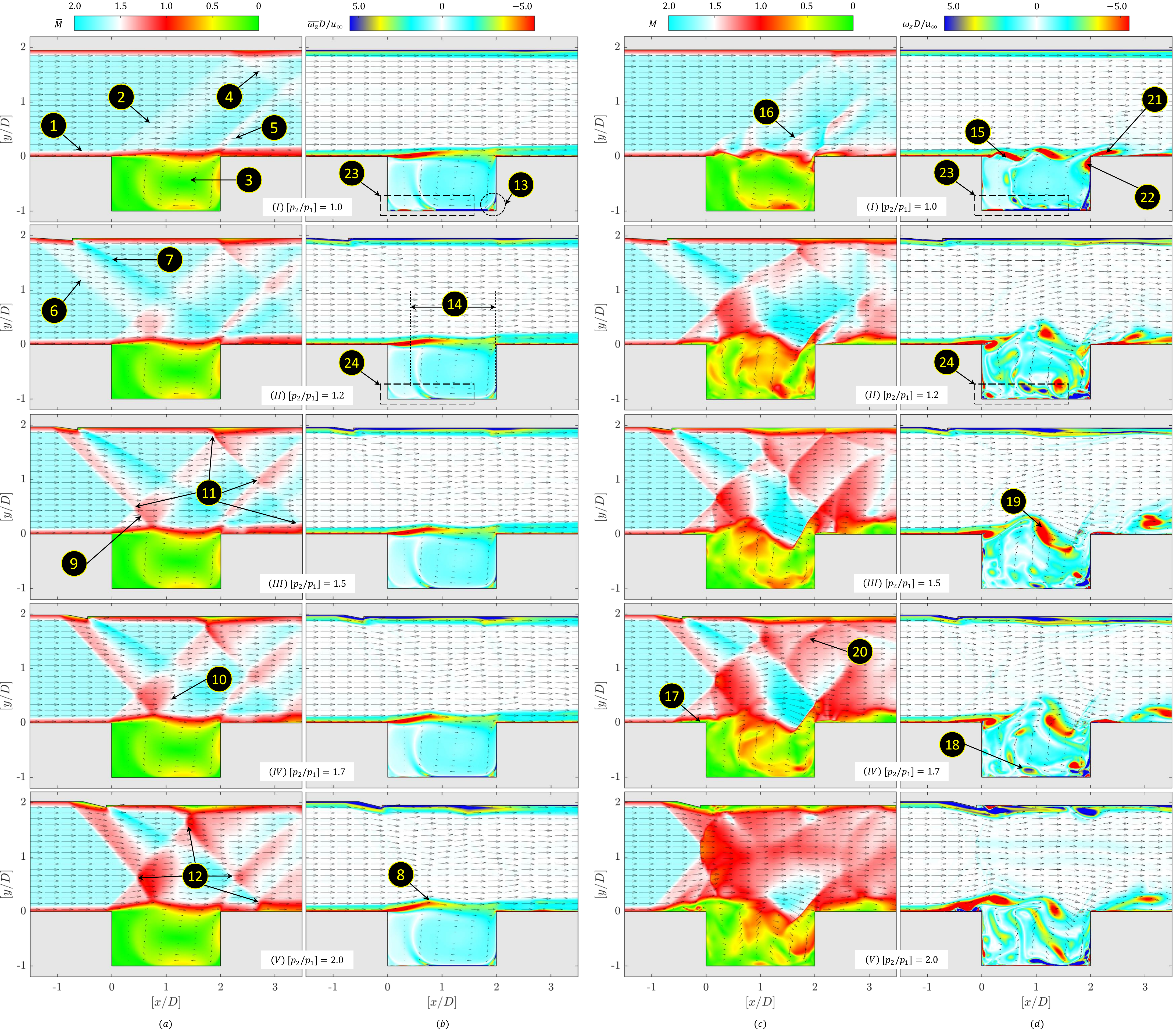}
	\caption{\label{fig:contour_plot} \sk{\href{https://youtu.be/nyLa3u4QoJ4}{(Multimedia View)} (a-b) Time-averaged and (c-d) instantaneous (at random time-step) contour plots of Mach number ($M$) and non-dimensionalized vorticity about the $z$-axis ($\omega_z D/u_\infty$) as in the left-hand coordinate system for different shock-shear layer interaction cases in a confined supersonic rectangular cavity flow: $(I-V)$ $[p_2/p_1]=$1.0, 1.2, 1.5, 1.7, and 2.0. Flow is from left to right. Distinct flow features: 1. incoming boundary layer, 2. shear layer induced compression wave, 3. cavity recirculation, 4. reflected-shock from the upper wall, 5. reattachment shock, 6. ramp-induced shock, 7. expansion at ramp-base, 8. shock-induced thickening of the cavity shear layer (visual shear layer thickening), 9. ramp-induced shock impinging on the shear layer after regular shock reflection, 10. impinging shock reflected as expansion wave by the shear layer, 11-12. regular/Mach type of shock interactions, 13. cavity-corner vortices, 14. length of the cavity recirculation region, 15. large-scale vortical structures, 16. traveling shocklets, 17. separated boundary layer, 18. vortical structures in the recirculation zone, 19. distorting vortical structures, 20. unsteady wavefront, 21-22. bifurcated vortical structure ejected to the freestream and trapped inside the cavity, and 23-24. strong and weak vorticity on the cavity-floor.}}
\end{figure*}

The solution file for each time-step from the two-dimensional DES simulation are exported as scattered data in ASCII (American Standard Code for Information Interchange) format and stored as a three-dimensional matrix for data processing. The scattered data points corresponding to the cavity wall and the two rakes (see Figure \ref{fig:prob_statement}) are extracted by interpolating them on a low-density grid for analysis purposes using a custom-made program in Matlab\textsuperscript{\tiny\textregistered} through a second-order interpolation scheme. \sk{As the interpolation is primarily carried out to convert the scattered solution data to a grid or uniformly spaced data of sufficient density, loss in spatial details are almost negligible.} The spatial data around the cavity for a certain extent ($-1.5 \leq [x/D] \leq 3.5$, and $-1.2 \leq [y/D] \leq 2.2$) are particularly interpolated on a two-dimensional structured coarse grid of equidistant points ($500 \times 500$) for representation purposes and also to perform modal analysis without the expense of huge computational power. A series of hybrid quiver-contour plots obtained in an aforementioned manner for each of the shock-shear layer interaction cases is shown in Figure \ref{fig:contour_plot}. The time-averaged spatial contours of the Mach number ($\overline{M}$) and the non-dimensionalized vorticity ($\overline{\omega_z} D/u_\infty$) are shown in Figure \ref{fig:contour_plot}a-b and the instantaneous contours for the same are shown in Figure \ref{fig:contour_plot}c-d. 

\begin{figure*}
	\includegraphics[width=0.88\textwidth]{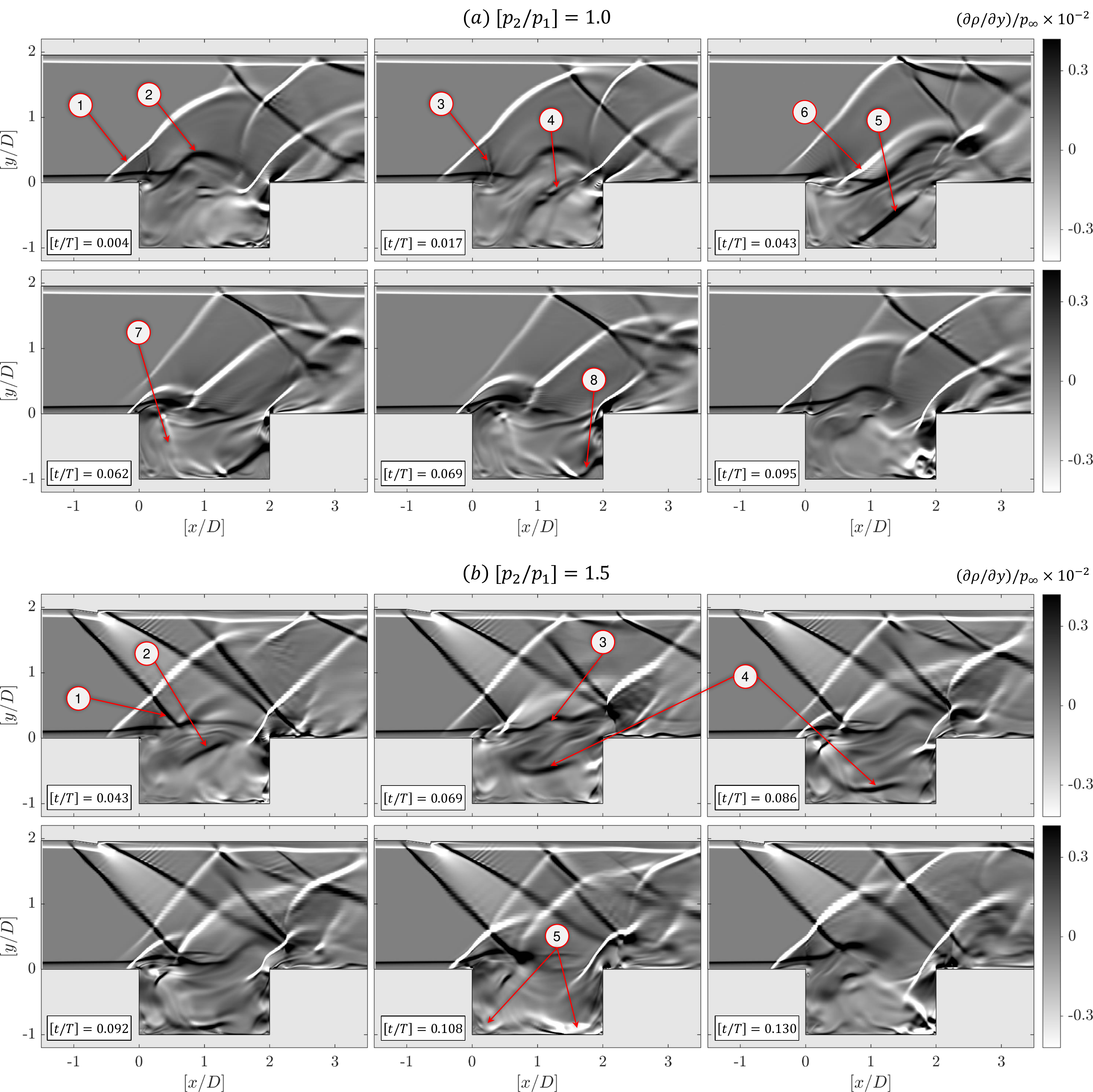}
	\caption{\label{fig:schlieren_snaps} \sk{\href{https://youtu.be/DHq-vqiDuoo}{(Multimedia View)} Snapshots taken at different non-correlated local time instants ($t/T$) showing the numerical schlieren along the $y$-direction for the two cases: (a) no shock-shear layer interaction ($p_2/p_1=1.0$) and (b) with shock-shear layer interaction ($p_2/p_1=1.5$). The snapshots explain the differences in the sound wave's emission inside the cavity from the shear layer. Flow features for (a): 1. separated shear layer induced shock, 2. distorting shear layer producing large-scale structures (KH instability), 3. shocklets produced due to shear layer distortion, 4. sound wave emitted from the distorting shear layer, 5. convecting sound wave focusing towards the front-wall corner, 6. longitudinal motion of reattachment shock in the supersonic stream, 7. reflected reattachment shock from the rear-wall inside the cavity, 8. collapse of a sound wave in the corner. Flow features for (b): 1. impinging ramp-generated shock on the shear layer, 2. transverse sound wave emission on the cavity floor by the impinging shock, 3. perturbed shear layer due to shock impingement, 4. radially spreading sound wave in the middle of cavity floor, 5. bifurcated disturbances running towards rear and front-wall.}}
\end{figure*}

Some of the key features like incoming boundary layer, separated shear layer induced shock, cavity recirculation, reflected-shock from the upper wall, and reattachment shock are seen in the no shock-shear layer interaction case (Figure \ref{fig:contour_plot}a-b, I). Features like the ramp-induced shock, expansion at ramp-base, shock-induced thickening of the cavity shear layer (visual shear layer thickening), impinging shock on the shear layer, impinging shock on the shear layer reflected as expansion wave, and regular/Mach type of shock interactions (upstream and downstream of the cavity) are seen in the other cases (Figure \ref{fig:contour_plot}a-b, II-V). The chaotic flow field at the time of shock-shear layer interaction is well illustrated in the instantaneous snapshot at an arbitrary $[t/T]$. Features like the shear layer vortical structures, traveling shocklets, separated boundary layer, vortical structures in the cavity recirculation zone, distorting vortical structures, and unsteady wavefronts are marked in Figure \ref{fig:contour_plot}c-d (\href{https://youtu.be/3hz0LZ9EBkQ}{Multimedia View}). Besides, a video showing the time-varying spatial contours of $[p/p_\infty]$ and $[\rho/\rho_\infty]$ for all the cases is also given in the supplementary under the name \href{https://youtu.be/5aa3lQNFo8A}{`video.mp4'}.

\subsection{Shock-shear layer interaction: Generic differences} \label{sec:generic_diff}

\sk{Before discussing the different shock-shear layer interaction cases, the generic differences between the clean and the shock-laden cavity flow are highlighted.} The contour plots in Figure \ref{fig:contour_plot} reveal some obvious events between the considered cases. As the ramp-shock strengthens, the shock-shear layer interaction becomes severe. The shock-induced mixing\cite{Rao2017} thickens the cavity shear layer between $1.2 \leq [p_2/p_1] \leq 2.0$ \sk{(see Figure \ref{fig:prob_statement}b and Table \ref{tab:case_list} for the definition of different cases)}. Regular and Mach type of shock reflections are observed at four different places in the considered fluid domain: 1. ramp-shock and separated shear layer induced shock interaction, 2. incident/reflected shock interaction with the upper-wall, 3. reflected shock and reattachment shock interaction, 4. reflected shock interacting with the lower-wall. The aforementioned regions have a regular shock reflection for $0 \leq [p_2/p_1] \leq 1.5$, whereas for higher strengths ($[p_2/p_1] > 1.5$) the reflections are predominantly of Mach type shock reflection. The shock angle between the impinging shock on the shear layer and the freestream differs based on the shock reflection types. The shock angle for Mach reflections is certainly higher than the shock angle for regular reflections. \sk{The ramp-generated shock-impingement point moves slightly upstream for Mach reflection case, and it eventually increases the recirculation region's length.} 

The shock impingement on the cavity shear layer alters the cavity dynamics significantly. Two cases are particularly considered to briefly highlight the differences: a. $[p_2/p_1]=1.0$ (clean cavity), and b. $[p_2/p_1]=1.5$ (shock-laden cavity). Numerical schlieren snapshots along the $y$-direction are shown in Figure \ref{fig:schlieren_snaps} at different local time instants for those two cases. When there is no shock-shear layer interaction, a sound wave or a shocklet from the distorted shear layer\cite{Kourta2001,Yang2019} convects along the freestream. As it convects, the wavefront is focused towards the front-wall corner. The wavefront bursts upon impingement and radiates sound everywhere inside the cavity. The upstream running wave further perturbs the shear layer (longitudinal motion). The perturbed shear layer is thus distorted again, producing another sound wave and keeps the feedback closed. 

However, during the shock interaction for cases like $[p_2/p_1=1.5]$, a cylindrical wavefront is generated when the shock impinges on the shear layer. The cylindrical wavefront impinges on the cavity floor's mid-section (transverse motion), particularly for the $[p_2/p_1]=1.5$ case. The wavefront bifurcates and travels to either side of the cavity (rear and front-wall), thereby perturbing the free-shear layer at the origin and the reattachment shock itself. The reflecting waves from the cavity wall have equal path differences and create resonances inside the cavity. On the contrary, for $[p_2/p_1]=2.0$ case, the impingement shock's upstream shift breaks the symmetry of the sound wave's impingement on the cavity-floor mid-section. Hence, the path difference between the reflected wave from the rear and front-wall will not be the same, and the cavity resonance vanishes. \sk{In the Figure \ref{fig:schlieren_snaps} \href{https://youtu.be/DHq-vqiDuoo}{(Multimedia View)}, the wavefront motion is shown through the $x-t$ diagram for a better understanding. A thorough discussion on the wavefront motion and the resonating cavity behavior for different cases is given further in the upcoming sections through detailed $x-t$ diagrams and spectral analysis.}

\subsection{Steady and unsteady statistics}\label{sec:steady_unsteady}

\begin{figure}
	\centering
	\includegraphics[width=\columnwidth]{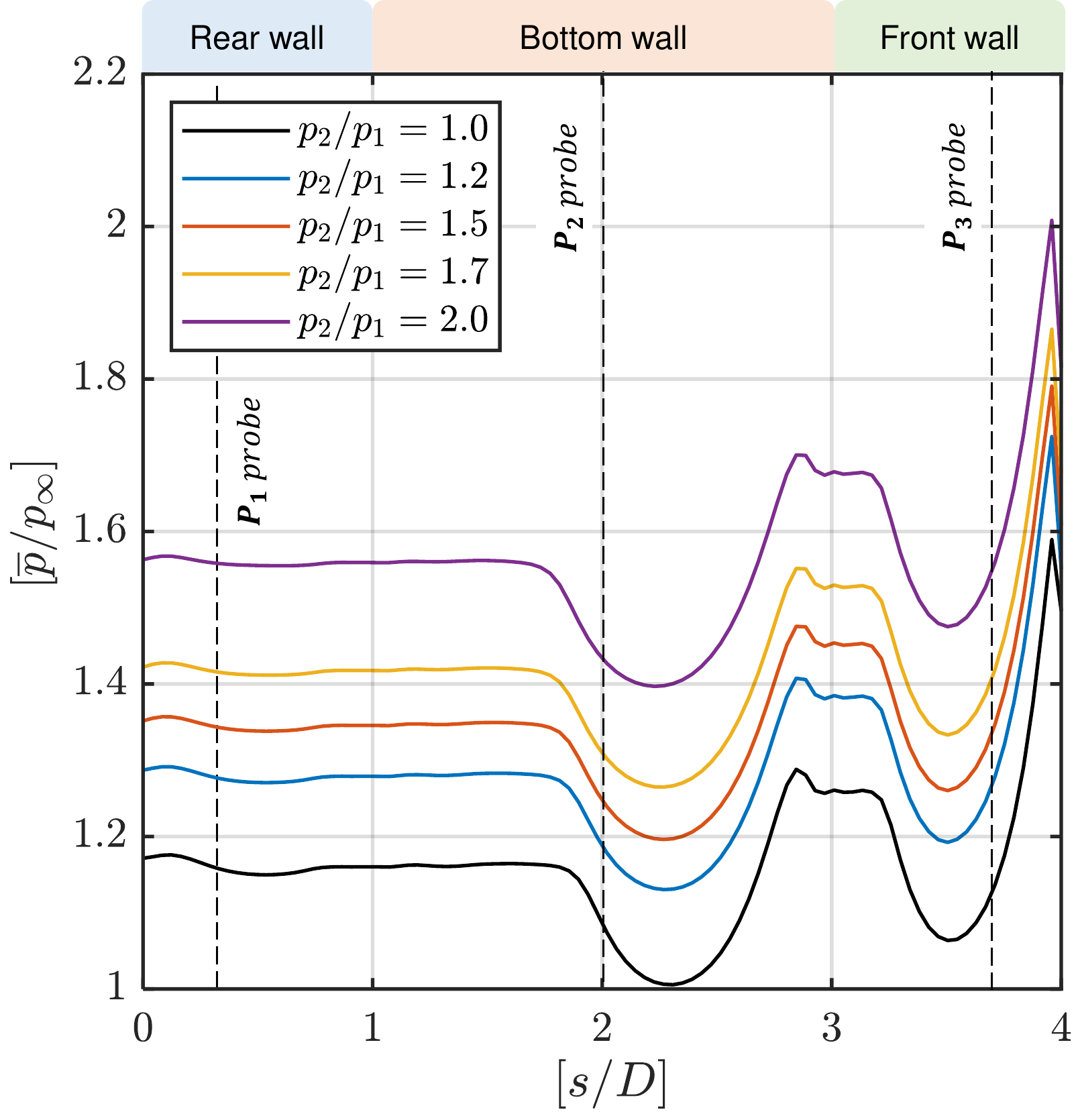}
	\caption{\label{fig:mean_pressure} Time-averaged wall-static pressure measurements ($\bar{p}/p_\infty$) from the RANS simulations for the different shock-shear layer interaction cases ($p_2/p_1=1.0,1.2,1.5,1.7,2.0$) along the surface of the cavity-wall ($s/D$) in a confined supersonic cavity flow.}
\end{figure}

Cavity wall-static pressure ($p/p_\infty$) is monitored to gather steady and unsteady statistics. Firstly, through the converged RANS simulation, time-averaged wall-static pressure ($\overline{p}/p_\infty$) is obtained for all the cases and is plotted in Figure \ref{fig:mean_pressure}. The peak-pressure for $[p_2/p_1]=1.0$ case is minimal and it is observed around the cavity-front wall edge ($s/D\sim 4$). The distorting vortical structures in the cavity shear layer impinge on the front wall, and the flow stagnates. The stagnated flow results in a huge pressure rise at $[s/D]\sim 4$. For $[p_2/p_1]>1.0$, the impinging shock on the cavity shear layer reflects as an expansion wave, as the shear layer is a constant pressure boundary. Due to the impingement of the shock wave on the shear layer, the shear layer thickens and produces disturbances that are comparatively larger in size for the canonical shear layer\cite{Shi2021}. Larger disturbances entrain higher fluid mass as it convects. Furthermore, the reflected expansion wave accelerates the flow downstream. The accelerated fluid mass trapped inside the large-scale structure impinges on the cavity front-wall and stagnates. The accelerated fluid mass stagnation results in further pressure rise at $[s/D]\sim 4$. In comparison to $[p_2/p_1]=1.0$, a maximum of 25\% rise in peak cavity wall pressure is seen for $[p_2/p_1]=2.0$ at $[s/D]\sim 4$. 

As the impinging shock strength increases ($p_2/p_1>1$), the peak pressure not only rises around the cavity's front-wall-edge but everywhere inside the cavity. After a critical value of $[p_2/p_1] \geq 1.5$, the Mach type of shock reflection increases the shock loading on the cavity, and the overall wall-static pressure increases, rapidly. \sk{The first bucket or drop-in $[\overline{p}/p_\infty]$ between $0 \leq [s/D] \leq 1$ marks the presence of corner vortices\cite{Biswas2020} (see label-13 in Figure \ref{fig:contour_plot}) that are conventionally seen in subsonic corner flows of lid-driven cavity flow. As $[p_2/p_1>1]$, the corner vortices disappear, and the suction level flattens. The dotted-boxed line drawn in Figure \ref{fig:contour_plot}b-I (label-23) sheds information about the disappearance. The positive vorticity existing on the cavity floor-wall indicates strong reverse flow for $[p_2/p_1 ]=1.0$. However, it is not the case for other shock-shear layer interactions. In-fact, the strength is so feeble that it is not visually identifiable in the present contour resolution. The bifurcated sound wave running to both left and right side of the cavity is attributed to the cavity's floor-wall vorticity changes as explained in $\S$\ref{sec:generic_diff}.} Another distinct feature that is evident from the graph of Figure \ref{fig:mean_pressure} is the second bucket zone between $1.5 \leq [s/D] \leq 3.0$. The suction experienced by the cavity bottom-wall is attributed to the recirculation zone formed inside the cavity. The length of the recirculation region gradually increases as $[p_2/p_1]>1$. The impingement of shock on the cavity-free shear layer causes enhanced fluid-mixing or shock-induced mixing\cite{Shi2021,Dupont2019}. The effectiveness of mixing increases with $[p_2/p_1]$ and so the recirculation region's length.

\begin{figure}
	\centering
	\includegraphics[width=\columnwidth]{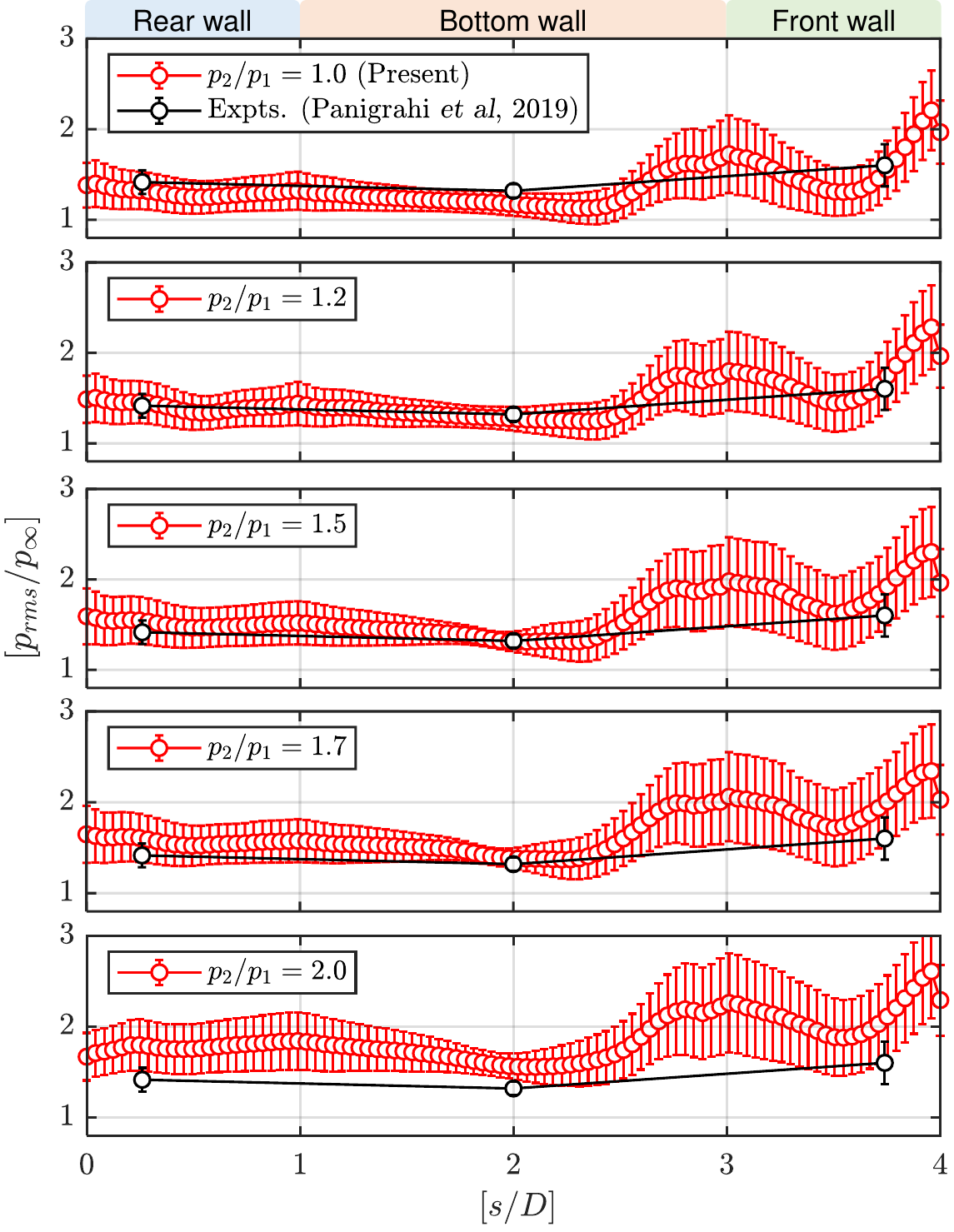}
	\caption{\label{fig:unsteady_press} Root-mean-square (\textit{rms}) variation of wall-static pressure along the surface of cavity with different shock-strengths ($[p_2/p_1]=1.0,1.2,1.5,1.7,2.0$) interacting with the cavity shear-layer. The markers represent the \textit{rms} and the error-bars represent the standard deviation of pressure fluctuations. Results are compared with the experiments of Panigrahi \etal \cite{Panigrahi2019} of similar flow conditions with no-ramp generated shock interacting with cavity shear-layer.}
\end{figure}

Unsteady statistics of wall-static pressure are extracted from the DES solution over a period of $2 \leq [t/T] \leq 4$ on the cavity-wall. The \textit{rms} (root-mean-square) values of static-pressure fluctuations ($p_{rms}/p_\infty$) are plotted in Figure \ref{fig:unsteady_press} and the fluctuation intensity or the standard deviation in the fluctuations is plotted as the error-bar. For all the cases, the fluctuations are intense in the lower part of the front wall or the corner. As the interacting shock strength increases ($p_2/p_1>1$), the fluctuation intensity increases drastically along with the \textit{rms} value. The aforementioned events can be reasoned using the video given in the supplementary (\href{https://youtu.be/5aa3lQNFo8A}{`video.mp4'}). The impinging shock on the cavity shear layer leaks through the vortical structures as cylindrical sound waves. These disturbances are radiated everywhere inside the cavity. However, the downstream convection of vortical structures focuses the wavefronts towards the front-wall corner. The part of fluid mass entrapped into the cavity from the previous impingement behind the reattachment shock is also pushed downwards to the front-wall corner. Similarly, the reflected cylindrical wavefronts from the rear-wall come back and interact with the front-wall corner. The cumulative effects of the focused wavefront entrapped fluid mass and the reflected rear-wall waves lead to the higher fluctuation intensity at $[s/D]=3$.

Unsteady observations of Panigrahi \etal \cite{Panigrahi2019} on the same cavity is also plotted against the present simulation for comparison. When there is no shock interaction, the experimental values are over-predicted, saying that the pressure experienced by the cavity wall in the experiments is relatively higher. As $[p_2/p_1]>1$, the cavity wall pressure in the computation increases. For $[p_2/p_1]=1.2$, the values match well with numerical findings. The respective experimental schlieren images (see Figure 6a-b in Panigrahi \etal \cite{Panigrahi2019}) reveal the presence of many impinging shocks, either from the rough surfaces or shock-induction by the large-scale structures in the upstream boundary layer. Thus, it is demonstrated in a quantified manner that the unsteady statistics are always higher than the predicted numerical counterpart in the actual experiments. The solvers' lower prediction is predominantly because the weak shock-shear layer interactions in the cavity are neglected by the smooth wall assumptions in the simulations. However, they are unavoidable in the experiments.

The fluctuation intensity helps monitor the changes in the cavity dynamics due to shock-shear layer interactions  (given by the error bar height in Figure \ref{fig:unsteady_press}). \sk{The fluctuation intensity variation between $0 \leq [s/D] \leq 4$ (rear, bottom, and front wall portion) follows a monotonically increasing trend about the mean as $[p_2/p_1]$ increases. However, the trend of $[p_{rms}/p_\infty]$ itself suddenly flips between $[p_2/p_1]=1.7$ and $2.0$.} Such behavior indicates the alteration of the shock reflections inside the cavity and the recirculation dynamics. In the next section, the shock-foot's motion on the cavity wall is monitored to get a clear picture.

\subsection{Cavity shock-footprint analysis}\label{sec:x-t}

\begin{figure*}
	\includegraphics[width=\textwidth]{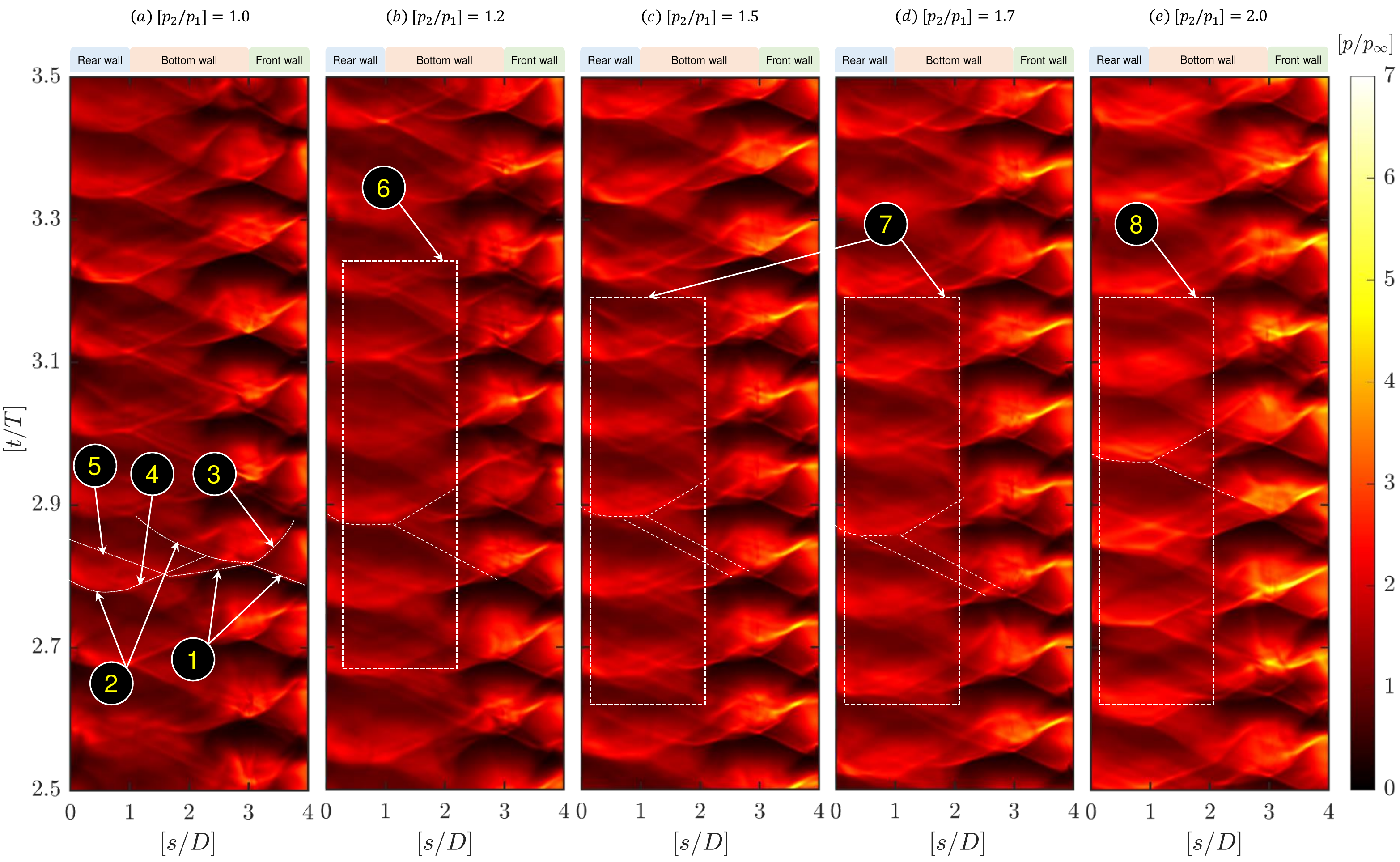}
	\caption{\label{fig:xt_press_cavity} \sk{(\href{https://youtu.be/_T6vZoj6-uA}{Multimedia View}) Wall-static pressure ($p/p_\infty$) trace over global flow time ($t/T$) along the cavity-wall surface ($s/D$) for different shock-shear layer interactions cases: (a) $[p_2/p_1]=1.0$, (b) $[p_2/p_1]=1.2$, (c) $[p_2/p_1]=1.5$, (d) $[p_2/p_1]=1.7$, and (e) $[p_2/p_1]=2.0$, in a confined supersonic rectangular cavity flow. Distinct features: 1. sound wave edge on the cavity-floor (right-running wave, RW) and front-wall (left-running wave, LW) converging towards the front-wall corner, 2. reflected disturbances from the sound wave impingement on the cavity-floor running towards (LW) the rear-wall corner and the leading edge, 3. disturbance running towards the cavity trailing edge after the sound wave implosion on the front-wall corner (RW), 4. reflected waves from the real-wall corner (RW), 5. refracted wave from sound wave impingement running towards the leading edge (LW). \sk{6-8. Different types of $x-t$ branch shapes (6. $90^\circ$ clockwise-rotated diffused-looking `Y'-type branches, 7. similar `Y'-type branch with an additional $x-t$ trace on the left-arm, 8. sharp looking `Y'-type branches)}. Corresponding video file showing the series of instantaneous pressure contours is also given separately in the supplementary (refer to \href{https://youtu.be/5aa3lQNFo8A}{`video.mp4'}).}}
\end{figure*}

The cavity dynamics are monitored by tracking the motion of pressure pulses on the cavity-wall by preparing the $x-t$ plot. A strong pulse marks the motion of shock-foot as a sharp line, and the slope of the line provides the disturbance's convection speed. The lines running towards the cavity leading/trailing edge are called left/right running waves (LW/RW). The wall surface coordinate system is considered (along the cavity-wall, $s/D$) and the wall-static pressure is grabbed for different time-steps (along the $y$-axis, $t/T$) to construct the $x-t$ diagram as shown in Figure \ref{fig:xt_press_cavity}. \sk{In the \href{https://youtu.be/_T6vZoj6-uA}{Multimedia View} of Figure \ref{fig:xt_press_cavity}, the process of $x-t$ diagram construction and the formation of $90^\circ$ clockwise-rotated `Y'-type branch is also visually demonstrated for a few unsteady cycle.} Some of the vital differences between the cases are easily highlighted in the $x-t$ diagram itself.

As $[p_2/p_1]$ increases, the plot brightness increases, especially in the front and rear wall, indicating the increasing effects of shock-loading (wall-static pressure rise) on the cavity. The dark patch found between $2 \leq [s/D] \leq 4$ for $p_2/p_1=1$ represents the largest suction or higher flow velocity. The higher flow velocity comes from the diversion of partially entrained fluid in the vortex pocket as it impinges the top sharp corner in the front-wall. For $[p_2/p_1]>1$, the vortex pocket-size increases due to the shock-induced mixing. The static pressure field closer to the front wall's reattachment shock region also increases due to multiple shock interactions. Eventually, further impingement of vortex pockets at higher $[p_2/p_1]$ lead to increased shock-loading on the front-wall. 

Close to the cavity's origin, the boundary layer detaches from the bottom wall and begins to curl-up and produce a rolling structure. The rolling structure with a stream of supersonic flow and subsonic flow on its side produces a shocklet. A part of the shocklet moves inside the cavity as a distinct sound wave, and the other travels in the supersonic stream (see also Figure \ref{fig:schlieren_snaps}). The rolling structures warp the shocklets as it convects. The sound wave travelling inside the cavity hits the front-wall and travels forth and back inside the cavity. The motion of these sound waves are shown in Figures \ref{fig:contour_plot}-\ref{fig:schlieren_snaps}. The LW between $3 \leq [s/D] \leq 4$ corresponds to the sound wave impinging on the front-wall, whereas the RW represents the reflected wave from the sound wave implosion on the front-wall corner. The RW further interacts with the wave system reflecting forth and back inside the cavity and refracts further. The `Y' -type branching is effectively seen at $[s/D] \sim 1$ and $[s/D] \sim 3$. As the wave system hits the rear-wall, it displaces the fluid from the cavity and perturbs the incoming shear-layer. The fluid displacement is very fast on the vertical rear-wall as the slope of the LW between $0 \leq [s/D] \leq 1$ is minimal. The ejection happening at the edge of the cavity shear layer helps maintain another vortex production, and thus the oscillation is self-sustained.

\begin{figure*}
	\includegraphics[width=\textwidth]{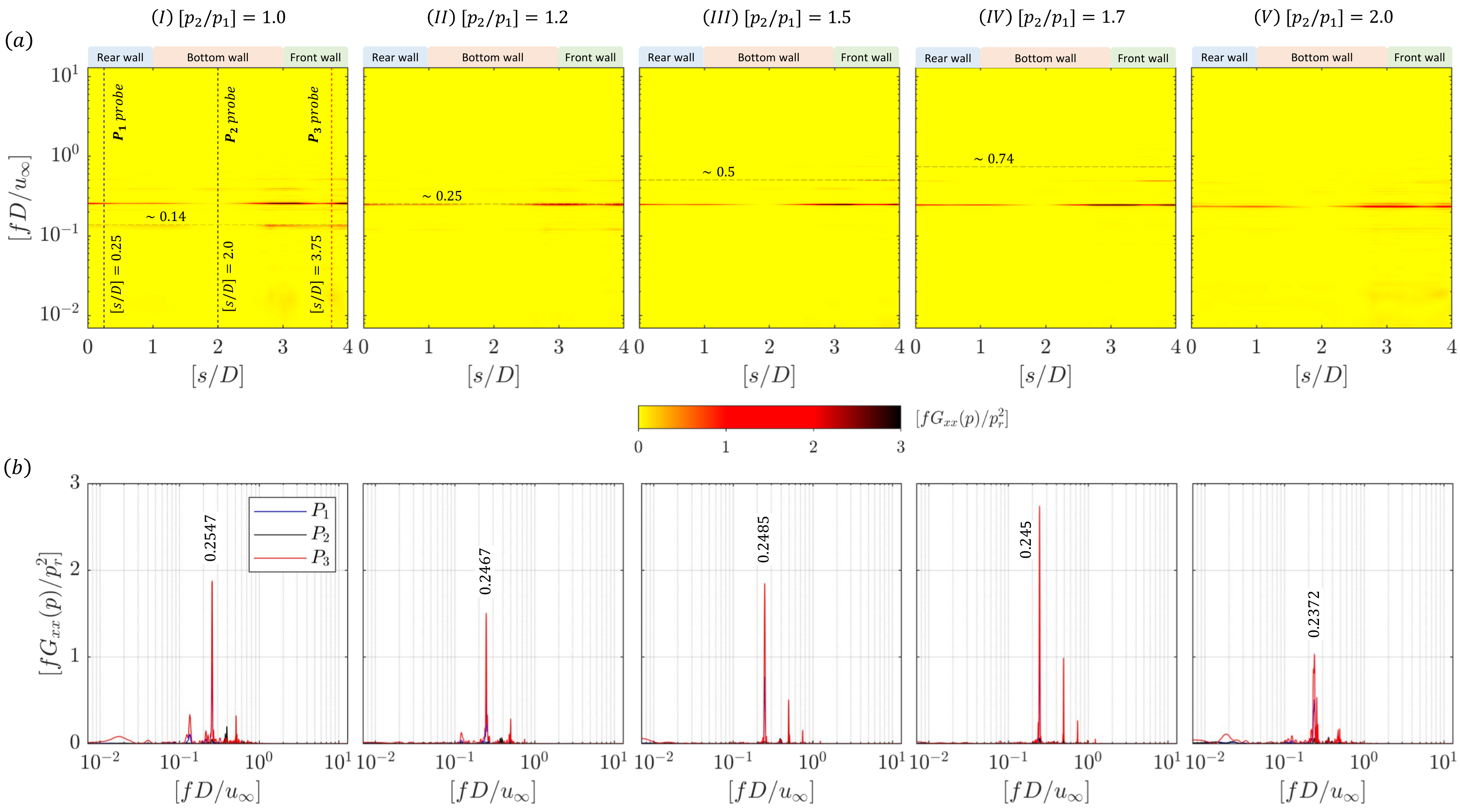}
	\caption{\label{fig:cavity_spectra} (a) Contour plots of non-dimensionalized power spectral density ($fG_{xx}(p)/p_r^2$, $p_r$=1 kPa) showing the dominant temporal modes observed from the wall-static pressure measurements ($p/p_\infty$) along the cavity surface ($s/D$) for shock-shear layer interactions of various shock-strength ($p_r$=1 kPa): (I) $[p_2/p_1]=1.0$, (II) $[p_2/p_1]=1.2$, (III) $[p_2/p_1]=1.5$, (IV) $[p_2/p_1]=1.7$, and (V) $[p_2/p_1]=2.0$, in a confined supersonic rectangular cavity flow; (b) Selective spectra at three probing locations ($P_1,P_2,P_3$) revealing the varying power and the occurrence of resonance frequencies for different shock-shear layer interaction strengths. Corresponding video file showing the series of instantaneous pressure contours is also given separately in the supplementary (refer to \href{https://youtu.be/5aa3lQNFo8A}{`video.mp4'}).}
\end{figure*}

Because of the shock interaction, the $x-t$ diagram produces more organized shock foot motion, in particular for $[p_2/p_1] = 1.5$ and $1.7$ cases. The periodicity in the shock-foot motion is consistent. The weaker waves are almost smudged in comparison to the sharp-lines representing the dominant shock-induced wave motion. Besides, due to the mid-section impingement of sound-waves from shock-shear layer interaction, there are two LWs seen for $[p_2/p_1] = 1.5$ and $1.7$ cases. The organized and comparatively more periodic features at higher shock-shear layer interaction case indicate a resonating behavior. However, as the shock interaction strength reaches $p_2/p_1=2$, the closely packed shock-foot motion disappears. Patterns of shock-foot of higher strength traveling at different speeds in comparison with the other cases are obvious. The branching points also change significantly, and the appearance of two LWs vanish. 

Two-point cross-correlation analysis is performed as shown in Figure \ref{fig:correlation_plots}a to identify the relation of longitudinal wave motion. The analysis reveals a distinct positive correlation at the cavity leading edge ($s/D=0$) to the front-wall corner ($s/D=3$) events. It represents the simultaneous upstream motion of the leading edge shock and the reattachment shock foot. However, closer to the trailing edge, the correlations are negative ($s/D\sim 4$). A pressure burst in the corner displaces fluid out of the cavity quickly and causes a pressure drop around the corner. The negative correlation behavior is attributed to the aforementioned event.

The constructed $x-t$ diagram is subjected to spectral analysis through Fast-Fourier transformation (FFT) to extract the dominant spectra at a given location. The respective non-dimensionalized power in the spectra is plotted as contours in Figure \ref{fig:cavity_spectra}a. Three selective locations are probed to monitor the power variations as shown in Figure \ref{fig:cavity_spectra}b. The probing locations are selected based on the unsteady pressure probe locations in the experiments of Panigrahi \etal \cite{Panigrahi2019}. When there is no shock-shear layer interaction, the Rossiter frequency corresponding to modes $n=1,2,3,4$ are visible at different power. When there is a shock-shear layer interaction, some of the modes vanish, and few of the modes are observed to be in resonance. At the time of resonance, harmonics in multiples of two are seen. At $[p_2/p_1=1.2]$, frequency corresponding to $n=1$ mode is very feeble. At $[p_2/p_1]=1.5$ and $1.7$, frequency corresponding to $n=1$ mode is vanished. However, a strong resonance is seen between the modes $n=2,4,6$ at $[p_2/p_1]=1.5$ and $1.7$. At $[p_2/p_1]=2.0$, no resonating behavior is seen, and comparatively broadband spectra centered around the mode $n=2$ is observed. Thus, for at least the considered cases, shock strength between $1.5 \leq [p_2/p_1] \leq 1.7$ is identified to be in resonance, where the generic Rossiter's frequency for a canonical cavity is tuned to particular modes.

In Figure \ref{fig:cavity_spectra}b, While monitoring the power specifically at $P_3$, a strong resonance power is observed at $[p_2/p_1]=1.7$ case ($\sim 45\%$ rise in $f\Delta G_{xx}(p)/p_r^2$ with respect to $p_2/p_1=1$) with a slight left-hand shift in the dominant spectra ($\sim4\%$ fall in $\Delta f D/u_\infty$ with respect to $p_2/p_1=1$). In the transition case ($p_2/p_1=1.2$), the power contents are observed to be smaller ($\sim 20\%$ rise in $f\Delta G_{xx}(p)/p_r^2$ with respect to $p_2/p_1=1$) than the no interaction case with a similar left-hand shift in the dominant spectra ($\sim2\%$ fall in $\Delta f D/u_\infty$ with respect to $p_2/p_1=1$). However, during the interaction of higher shock strength with the cavity shear layer ($p_2/p_1=2$), the power contents drop significantly ($\sim 47\%$ drop in $f\Delta G_{xx}(p)/p_r^2$ with respect to $p_2/p_1=1$) and a prominent shift in the dominant broadened spectra to the left-hand side ($\sim7\%$ fall in $\Delta f D/u_\infty$ with respect to $p_2/p_1=1$) is seen. The observation of periodicity or patterns in $[p_2/p_1]=1.5$ and $1.7$ cases in Figure \ref{fig:xt_press_cavity} are thus found to be in accordance with the spectral analysis shown in Figure \ref{fig:cavity_spectra}. Similarly, a new pattern of different shock-foot print strength seen in $[p_2/p_1]=2.0$ case is also in agreement with the completely altered spectra in Figure \ref{fig:cavity_spectra}. 

\subsection{Shock oscillation in the duct}\label{sec:shock_osci_duct}

\begin{figure*}
	\includegraphics[width=\textwidth]{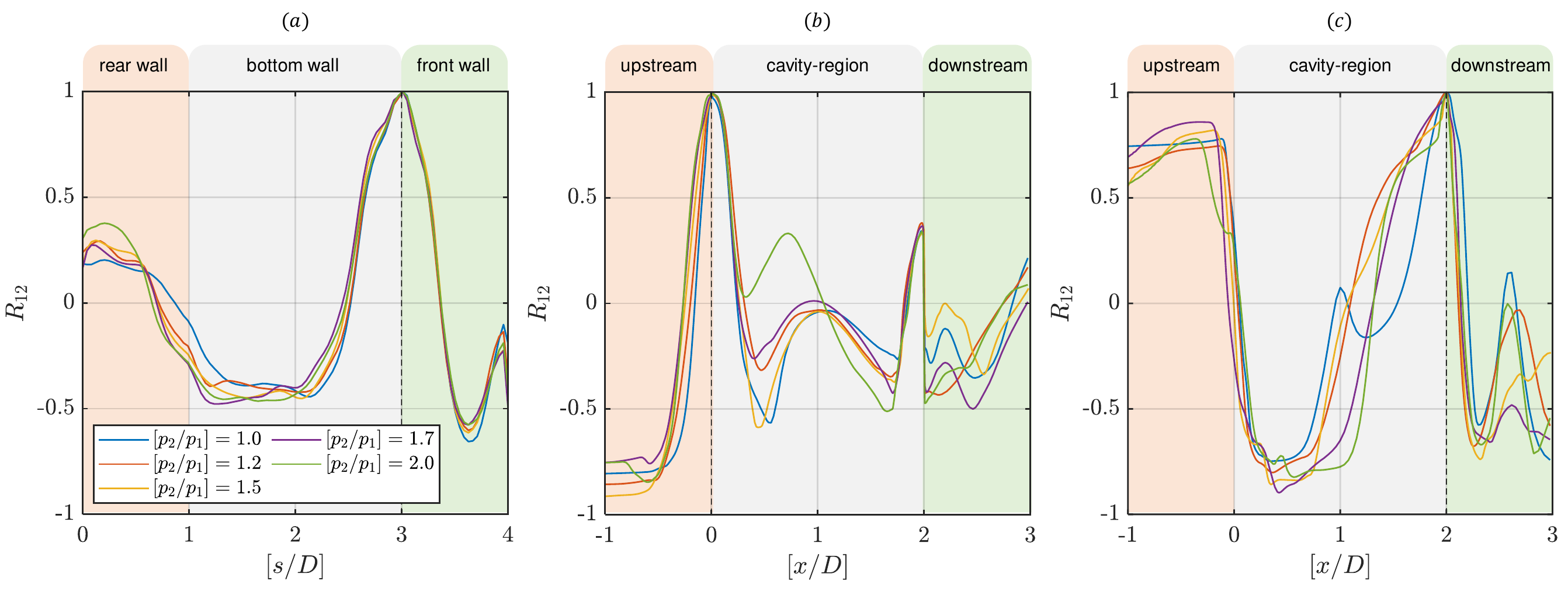}
	\caption{\label{fig:correlation_plots} Two-point spatial correlations ($R_{12}$) observed for the static pressure fluctuations ($p/p_\infty$) in the $x-t$ diagrams of Figures \ref{fig:xt_press_cavity} and \ref{fig:xt_press_rake}: (a) along the cavity wall surface with respect to $[s/D]=3$), (b) along the probing rake at $[y/D]=0.1$ with respect to $[x/D]=0$, and (c) along the probing rake at $[y/D]=0.5$ with respect to $[x/D]=2$.}
\end{figure*}

\begin{figure*}
    	\includegraphics[width=\textwidth]{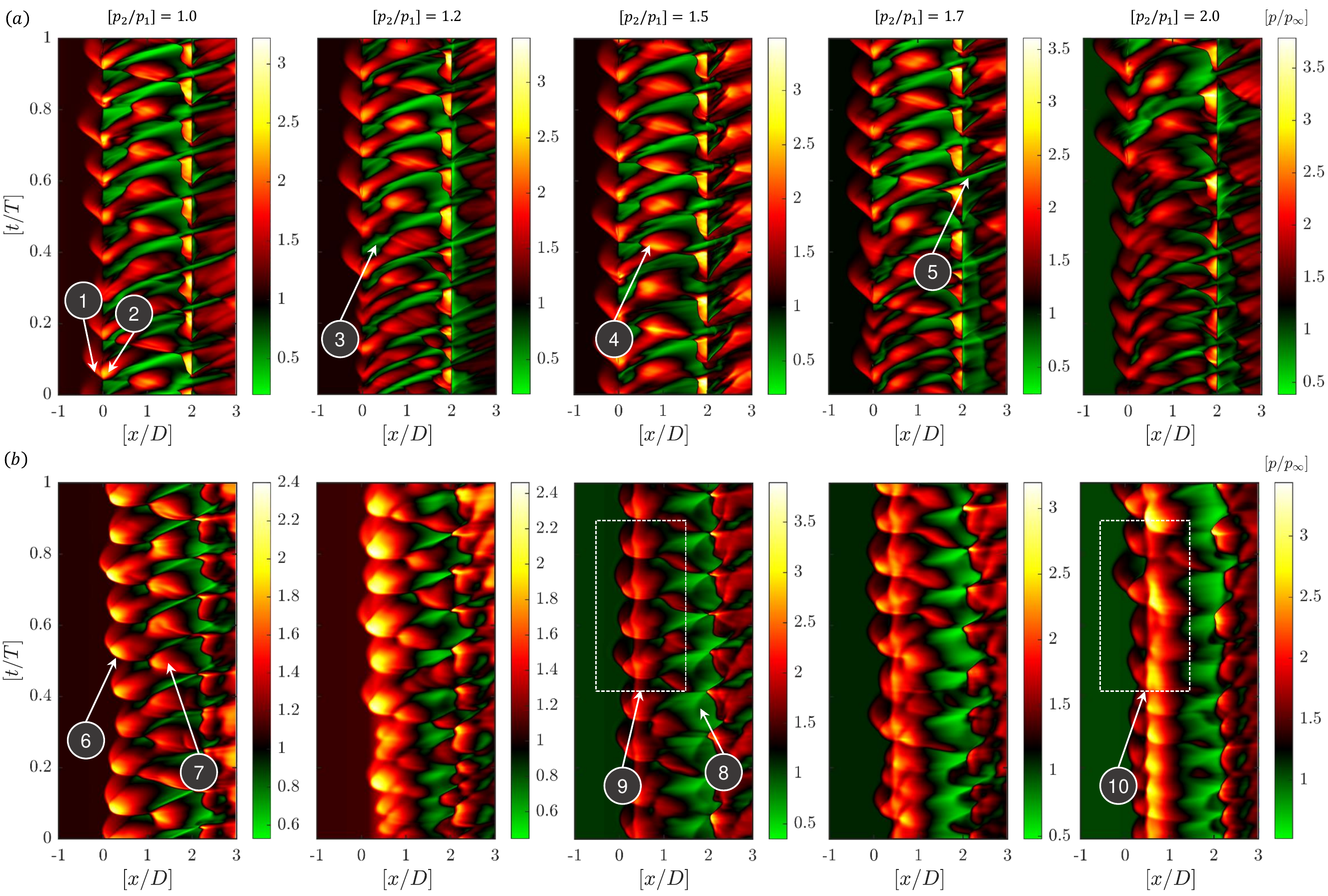}
	\caption{\label{fig:xt_press_rake} \sk{\href{https://youtu.be/jMkK06uD-lU}{(Multimedia View)} $x-t$ diagram constructed from the non-dimensionalized static pressure distribution ($p/p_\infty$) at different local time instants ($t/T$, $T=1$ ms) for different shock-shear layer interaction cases along the two different rake locations: (a) $[y/D]=0.1$ and (b) $[y/D]=0.5$ in a confined supersonic rectangular cavity flow. Distinct features: 1. perturbed shock moving upstream (bump-like structures), 2. shocklet from the vortex convecting downstream, 3. convecting vortical structure (minimum pressure in the core), 4 \& 7. longitudinal motion of reattachment shock, 5. partially expelled vortical structure to the freestream, 6. bubble-like structures formed due to the leading edge shock perturbation, 8. expansion-fan from the top-wall ramp, 9. organized motion of the shocks inside the duct, 10. chaotic motion of the shocks inside the duct. Corresponding video file of the instantaneous pressure contours is also given separately in the supplementary (refer to \href{https://youtu.be/5aa3lQNFo8A}{`video.mp4'})}}
\end{figure*}

\begin{figure*}
	\includegraphics[width=\textwidth]{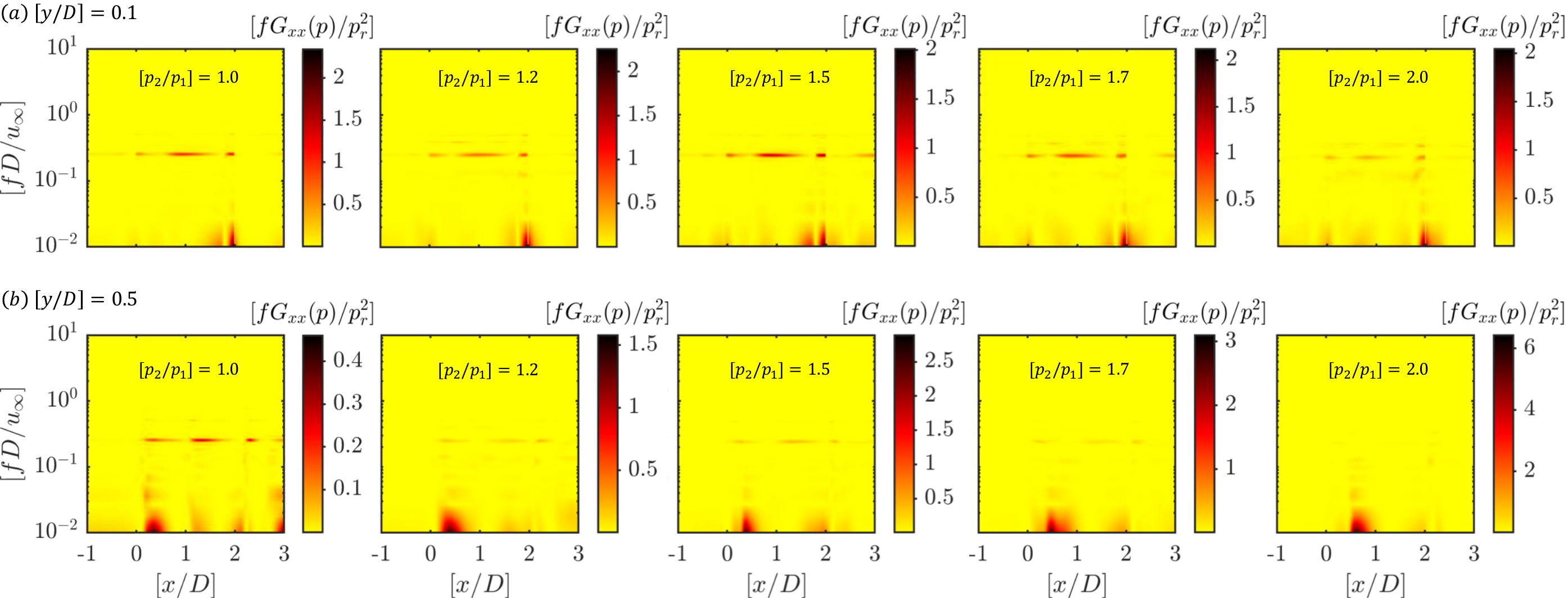}
	\caption{\label{fig:rake_spectra} Contour plots of non-dimensionalized power spectral density ($fG_{xx}(p)/p_r^2$, $p_r$=1 kPa) showing the dominant temporal modes observed from the static pressure measurements ($p/p_\infty$) at two different rake locations: (a) $[y/D]=0.1$ and (b) $[y/D]=0.5$ for shock-shear layer interactions of various shock-strength in a confined supersonic rectangular cavity flow. Corresponding video file showing the series of instantaneous pressure contours is also given separately in the supplementary (refer to \href{https://youtu.be/5aa3lQNFo8A}{`video.mp4'}).}
\end{figure*}

The impinging shock from the ramp alters not only the dynamics of the cavity but also the overall dynamics closer to the duct's exit. The impinging shock after interacting with the compression shock from the shear layer reflects as an expansion wave. The reflected expansion wave from the free shear layer oscillates along with the shedding of vortical structures. The upstream shock structure oscillations thus influence the downstream shock reflections. The coupling or coherence in the shock oscillations can once again be studied using the $x-t$ diagram as shown in Figure \ref{fig:xt_press_rake} by monitoring the shock motion along with different streamwise rake locations away from the cavity-floor. Such a study will help use the shock-laden flow stream as a passive control jet discussed separately in the upcoming section. 

In Figure \ref{fig:xt_press_rake}, $x-t$ contour plots are shown for two different locations: (a) closer to the cavity ($y/D=0.1$) and (b) slightly away from the cavity ($y/D=0.5$). \sk{The \href{https://youtu.be/jMkK06uD-lU}{Multimedia View} given in Figure \ref{fig:xt_press_rake} also demonstrate the $x-t$ plot construction procedure visually at those two locations for a few unsteady cycle.} For plots drawn at $[y/D]=0.1$ (Figure \ref{fig:xt_press_rake}a), some of the following features are evident at least in the $[p_2/p_1]=1.0$ case: the repeated forcing of the incoming boundary layer to separate (seen by the bumps formed between $-1 \leq [x/D] \leq 0$), the convection of shocklets or vortices from the cavity leading edge to the trailing edge (seen as streaks of alternating green and red lines between $0 \leq [x/D] \leq 2$), and the ejection of fluid mass from the cavity along with the incoming vortex pocket (seen as sharp green streak spaced between a wide region of a red patch at $2 \leq [x/D] \leq 3$). While monitoring the $x-t$ plot, the phase of the emanating leading-edge pressure pulse (in the form of bump-like structure) and the phase of the ejected fluid mass in the cavity trailing edge (in the form of the green streak) are observed to be phase-locked (happens at the same time). In the two-point cross-correlation plot of Figure \ref{fig:correlation_plots}b, the upstream shock in the leading edge ($x/D=0$) and the region around the reattachment shock ($x/D\sim 2$) are also shown to be positively correlated, confirming the phase-locking behavior.

In the $x-t$ plots drawn at $[y/D]=0.5$ (Figure \ref{fig:xt_press_rake}b), different features are seen for the $[p_2/p_1]=1.0$ case. Repeated bubble-like structures in reddish-orange color emanate along the temporal direction at $[x/D]\sim 1$. As the interacting shock strength increases, for $[p_2/p_1]=1.2$, the ramp's incident shock and the compression shock from the shear layer interact periodically. It results in regular shock reflection, and a chain of downstream shock interactions is visible. The pattern looks more complicated than the $[p_2/p_1]=1.0$ case. However, the pattern's organization from the shock oscillation is periodic only for the $[p_2/p_1]=1.5$ case. Consistent shock reflections are seen between $0 \leq [x/D] \leq 3$ and the trace of reflected expansion fan (green in color) is seen. For $[p_2/p_1]=1.7$, the patterns are distorted as time progresses. For $[p_2/p_1]=2.0$, the pattern is completely different, although it looks periodic to a certain temporal region. The last case contains severe shock-shock interaction inside the duct as the rake leading-edge is dominated by the Mach reflections (yellow bubble-like structures between $0 \leq [x/D] \leq 1$). The convecting waves with the shear layer structures and the expansion fan's associated motion are also not periodic. The other shock systems severely intercept them. From the video given in the supplementary (refer to \href{https://youtu.be/5aa3lQNFo8A}{`video.mp4'}) and also from the \href{https://youtu.be/nyLa3u4QoJ4}{Multimedia View} of Figure \ref{fig:contour_plot}, the shock intercepting behavior is seen. 

In the two-point spatial cross-correlation plot, as shown in Figure \ref{fig:correlation_plots}c, the influence of impinging shock is clearer. In the region between $1\leq [x/D] \leq 2$, the correlation switches immediately when there is a shock interaction. The chain of upstream and downstream shock/expansion wave motion concerning the reattachment shock displacement is identified to be the reason. However, downstream the cavity's trailing edge ($x/D \sim 2.5$), the correlations are negative only for particular cases ($p_2/p_1=1.5$ and $1.7$). The periodical fluid mass displacement in response to the reattachment shock motion from the cavity resonance is attributed to the aforementioned behavior. As the cavity resonance breaks at high or low impinging shock strength ($p_2/p_1=1.2$ and $2.0$), there is no correlation at $[x/D] \sim 2.5$, just like the no shock-shear layer interaction case ($p_2/p_1=1.0$).

The dominant spectral contents inside the confined duct are accessed through the $x-t$ diagram plotted in Figure \ref{fig:xt_press_rake}. FFT analysis is performed at every spatial point, and a non-dimensionalized power spectral density contour plot is drawn as shown in Figure \ref{fig:rake_spectra}. At $[y/D]=0.1$, the spectra is almost similar to that of the cavity wall spectra shown in Figure \ref{fig:cavity_spectra}. However, few features are different in Figure \ref{fig:rake_spectra}a. The dominant spectra is slightly broadened, and there is a low-frequency component seen at the cavity trailing-edge (around $x/D \sim 2$ with $fD/u_\infty\sim 10^{-2}$). The low frequency is attributed to the ejected mass out of the cavity as seen from the multimedia-view in Figure \ref{fig:contour_plot}. As $[p_2/p_1]$ increases, the intensity of ejection rate increases and also looks broadened. Discrete frequencies corresponding to Rossiter's frequency of $n=2$ and $4$ are dominant and appears to be resonating for $[p_2/p_1]=1.5$. For $[p_2/p_1]=1.2, 1.5$ and $1.7$, along with the frequency corresponding to $n=2$ mode, there are other harmonics weakly present. At $[p_2/p_1]=2$, the power content is almost completely shifted to the low frequency contents.

Performing a spectral analysis away from the cavity at $[y/D]=0.5$ (Figure \ref{fig:rake_spectra}b) reveals other interesting features. The leading edge pressure pulse emission from the cavity displaces the boundary layer and separates as shown in Figure \ref{fig:contour_plot}. The resulting displacement sends a strong sound pulse into the supersonic freestream (zone of action) relatively. The resulting shock motion is felt all the way to the middle of the duct until the shock-reflection readjusts to the downstream condition. Hence, for all the cases, there is a strong low-frequency component felt at $[x/D] \sim 0.5$ with $[fD/u_\infty]\sim 10^{-2}$. The aforementioned component is broadened except at $[p_2/p_1]=1.5$ and $1.7$ where the shock-system resonates with the cavity. When $[p_2/p_1]$ increases, the power contained in the dominant frequency decreases initially for $[p_2/p_2]=1.2$ and moderately increases for $[p_2/p_1]=1.5$. The strength weakens immediately for $[p_2/p_1]=2.0$, and the dominant frequency corresponds to $n=2$ mode vanishes. Once again, the change of shock reflection pattern from regular to Mach breaking the resonance between the shock and cavity oscillation is attributed to the aforementioned behavior.

\subsection{Dominant temporal modes}\label{sec:modes}

\begin{figure*}
	\includegraphics[width=\textwidth]{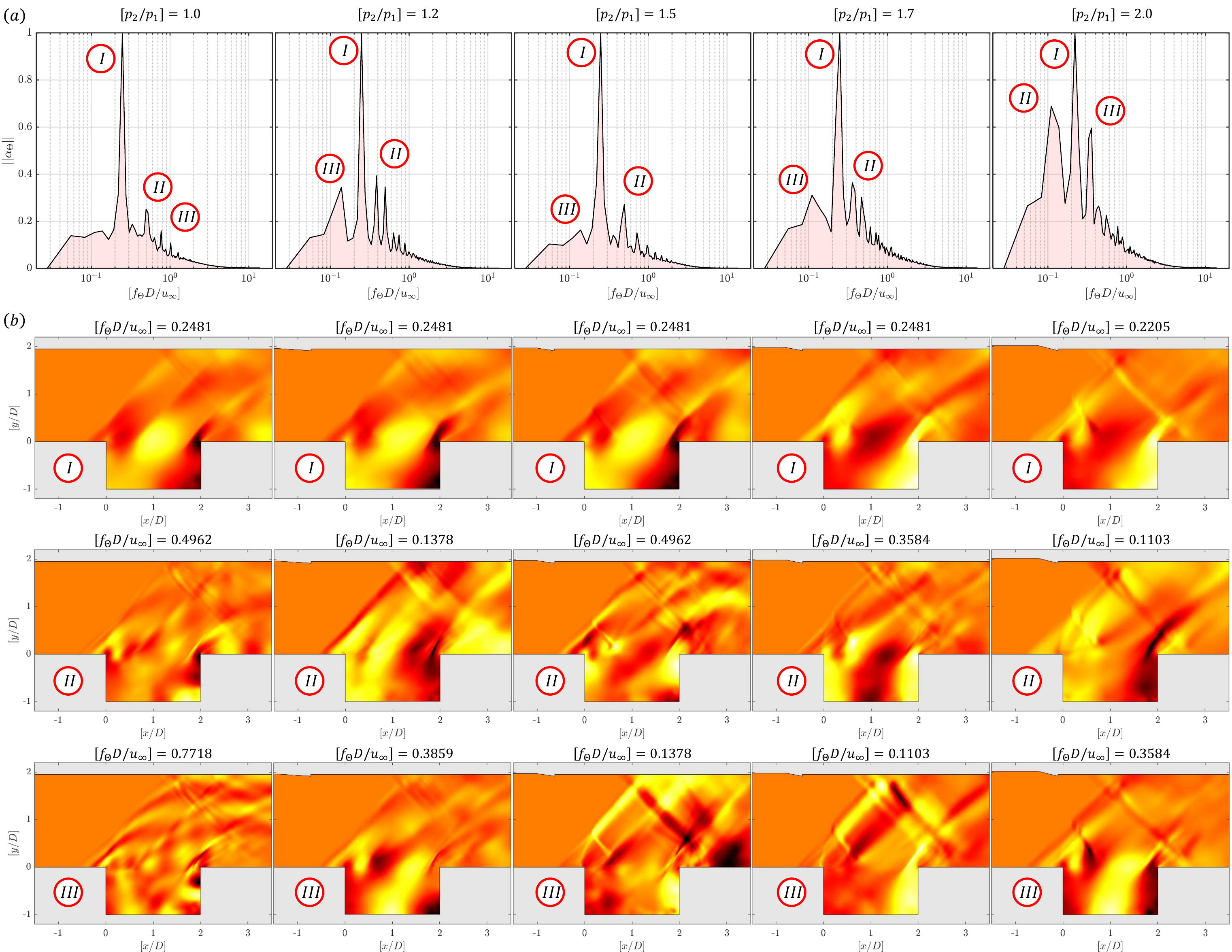}
	\caption{\label{fig:dmd_modes} (a) Typical normalized temporal spectra from the DMD analysis ($f_\Theta$ vs. $||\alpha_\Theta||$) for each case of the shock-shear layer interaction in a confined supersonic rectangular cavity flows. (b) Typical normalized spatial modes ($\Theta\left(x/D,y/D\right)$) corresponding to the unsteady static pressure variations based on the spectral content ($f_\Theta$) of decreasing amplitudes (marked as I, II, and III) in a confined supersonic rectangular cavity flows. The values of color contour (black-red-yellow-white) are between -1 to 1. Corresponding video file showing the series of instantaneous pressure and density contours is also given separately in the supplementary (Refer \href{https://youtu.be/5aa3lQNFo8A}{`video.mp4'}).}
\end{figure*}

The time dynamics of the static pressure in the entire spatial domain are analyzed for all the different cases through a modal analysis scheme called DMD\cite{kutz,taira,Rao2019a,Sahoo2021} (dynamic mode decomposition). The resulting analysis reveals the spatial modes that are temporally orthogonal. The time-dynamics involving the entire spatial field are computed first. Based on the largest amplitude in the individual spectra, the corresponding temporal mode is investigated for every case. The steps involved in the DMD routines are fairly straight forward and hence, they are not discussed here for brevity. Spatial data between $2\leq [t/T] \leq 1$ are collected at $[\Delta t/T]=1\times 10^{-3}$ interval. A column matrix is constructed only with the spatial pressure fluctuations, and the decomposition is performed.

In Figure \ref{fig:dmd_modes}a, the global spectra and the normalized amplitude is given for each of the cases. Initially at $[p_2/p_1]=1$, there is only one dominant frequency corresponding to $n=2$. When the shock interaction strength increases gradually, other harmonics of higher/lower amplitudes are seen. For $[p_2/p_1]=1.2$, frequencies corresponding to mode numbers $n=1,2,3,4$ are visible. At $[p_2/p_1]=1.5$, resonating modes $n=2$ and $4$ are only present. For $[p_2/p_1]=1.7$, the spectra become broadened, however the harmonics are seen to be with lower amplitude. At $[p_2/p_1]=2.0$, the harmonics shift slightly left and gain amplitude significantly, especially for frequencies closer to Rossiter mode $n=1,2$ and $3$. 

Spatial modes correspond to the first three dominant frequencies for each of the cases are plotted in Figure \ref{fig:dmd_modes}b. Spatial modes corresponding to $n=2$ are identified to be the dominant ones in all the cases. In a simplified sense, the contours can be interpreted as spatial correlations varying between -1 to 1 (black-red-yellow-white). The alternate shedding of large coherent structures of at least three numbers are seen between $0 \leq [x/D] \leq 2$ for the cases of $1 \leq [p_2/p_1] \leq 1.5$. For $[p_2/p_1]>1.7$, the impinging shock and the shock-induced mixing produces four to five small coherent structures. The resulting spatial mode looks completely different than the no-shock interaction case. Thus, it can be concluded that a shock strength of up to $[p_2/p_1]=1.5$ is permitted to attain the dominant spatial field as equivalent to the no-shock interaction case. Any shock interaction of strengths higher than $[p_2/p_1]>1.5$ results in the different dominant spatial fields, at least in representing the structures in the cavity shear layer.

\begin{figure*}
	\includegraphics[width=\textwidth]{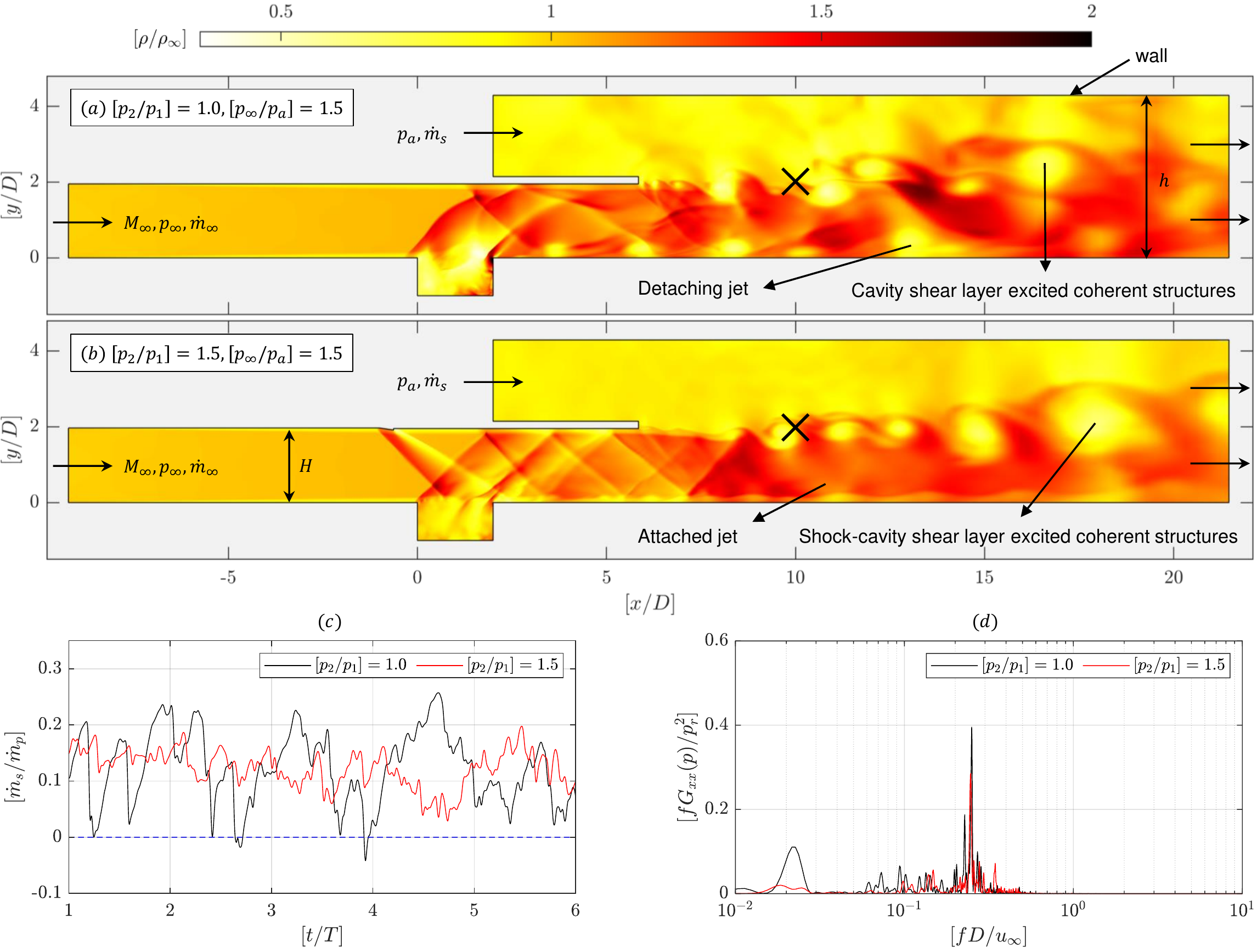}
	\caption{\label{fig:control_case} (\href{https://youtu.be/0xYP3NJdMhw}{Multimedia View}) Instantaneous snapshot showing the non-dimensionalized density field contours ($\rho_2/\rho_\infty$) at an arbitrary non-correlated time-step in a confined supersonic wall-jet flow: (a) without $(p_2/p_1=1.0)$ and (b) with $(p_2/p_1=1.5)$ shock-shear layer interaction in the cavity. (c) Temporal variation in the entrainment ratio ($\dot{m}_s/\dot{m}_p$) between the two cases showing damped oscillations for $[p_2/p_1]=1.5$. (d) Non-dimensionalized power spectral density ($fG_{xx}(p)/p_r^2$, $p_r=1$ kPa) of static pressure fluctuations probed at a particular location along the jet shear layer ([$x/D,y/D]=[10,2]$, shown as `$\times$' in Figure \ref{fig:control_case}a-b) between the two cases showing the tuned frequency for $[p_2/p_1]=1.5$. Flow conditions: $M_\infty=1.71$, $[p_a/p_\infty]\approx 1.5$, and $\overline{\Omega} = [\overline{\dot{m}_s}/\overline{\dot{m}_p}]\approx 0.14$ (Note: For the same $\overline{\Omega}$, in case-(a), $\Omega_\sigma = 0.0656$, and in case-(b), $\Omega_\sigma = 0.0275$. A reduction of about 58\% in the fluctuation intensity of $\Omega$ is seen for case-(b)).}
\end{figure*}

The spatial mode corresponding to higher harmonics varies almost for all the cases. For $[p_2/p_1]=1.0$, the harmonics are weak, and the respective spatial modes contain only fine structures. For $[p_2/p_1]=1.5$, the motion of shocks inside the duct and the shock-foot in the cavity are strongly correlated, reassuring the coupling behavior. However, the motions are anti-correlated for the third dominant tone, representing the out-of-phase or phase lag. The influence of impinging shock is not at all prominent in the first three dominant modes for $[p_2/p_1]=1.0$ and $1.2$. Once the shock strength increases to $[p_2/p_1]=1.5$, the influence of impinging shock is strongly felt in the second and third spatial modes. At $[p_2/p_1]=1.7$, a similar behavior is seen at a slightly different frequency. For $[p_2/p_1]=2.0$, the shock systems are visible in the first mode itself. A strong correlation between the duct shock and the wavefronts inside the cavity is seen at other spatial modes. The extent of upstream boundary layer displacement is also comparatively larger in comparison with the other cases. The findings are following the results of spectral and two-point cross-correlation analysis.

\section{Shock-laden cavity as a passive control device} \label{sec:application}

The present analysis demonstrates that the cavity shock-shear layer interaction of particular strength (here it is $p_2/p_1=1.5$) tunes the cavity to produce a discrete pulse corresponding to Rossiter's mode of $n=2$ and $4$. Both the cavity and the shock-system inside the duct are also coupled and resonating at significant amplitude. If the shock oscillations contain discrete frequencies in shock-laden duct flow problems, then the exhausting jet will be periodically forced. A periodically forced free-jet exhibits enhanced flow mixing and reduced far-field noise emission\cite{Rao2020b,Baskaran2019}. In a confined jet flow studies\cite{Karthick2016}, due to jet-column instability, certain over-expanded jets screech and produces huge noise. Although they are beneficial in terms of flow mixing, the aeroacoustic loads are severe\cite{Krothapalli1996}. In these cases, the results of the shock oscillations in a confined supersonic cavity flow with shock-shear layer interaction can be used to engineer a flow-control device. 

A computational study is performed to demonstrate the flow-control ability of a shock-laden cavity, particularly to the confined supersonic jets. A planar supersonic wall-jet case is considered where the jet exhaust is shrouded by a wall present at a height of $[H/h]\approx 2$. The inlet of the shroud is open to the ambient, and the geometrical details are given in Figure \ref{fig:control_case}a-b. The primary jet Mach number is kept at $M_\infty=1.71$, and the static pressure ratio between the primary jet and the ambient is kept at $[p_\infty/p_a]\approx 1.5$. Due to the high-speed jet (primary flow) exhaust, a secondary flow is inducted into the confinement as seen in a typical supersonic ejector\cite{Rao2014}. Two cases are run using the same DES computational methods mentioned in $\S$\ref{sec:numerical}: a cavity (a) without and (b) with shock interaction ($p_2/p_1=1.5$) in the primary flow.

The non-dimensionalized density contours ($\rho/\rho_\infty$) between the two cases are plotted first in Figure \ref{fig:control_case}a-b to understand the basic flow pattern. In the first case, the jet-column instability is predominant, and the jet disintegrates quickly. The cavity excited large-scale structures are visible. However, in the shock-laden cavity case, the jet-column is comparatively stable and attached to the bottom wall. Shock-laden cavity-excited large-scale structures are well-organized in comparison with the other case. The ratio of secondary and primary mass flow is called as entrainment ratio ($\Omega=\dot{m}_s\dot{m}_p$), which is considered as an important parameter to evaluate the performance of confined jet flows. Time-varying entrainment ratio ($\Omega(t/T)$) is plotted in Figure \ref{fig:control_case}c for both cases. Owing to the jet-column instabilities, the first case's entrainment contains huge fluctuations ($\Omega_\sigma=6.6\%$) about the mean. However, in the control case (shock-laden cavity flow), the fluctuations are minimal ($\Omega_\sigma=2.8\%$) about the mean. A significant reduction of about 58\% on the entrainment ratio fluctuations is seen between the two cases. It has to be noted that the average entrainment ratio between the two cases remains the same at $\overline{\Omega}\approx 0.14$.

The cavity tunability is explained by probing the static pressure at a point $[x/D,y/D]=[10,2]$ on the jet shear layer between the two cases. Spectral analysis results from the pressure fluctuation signals are shown in Figure \ref{fig:control_case}d. When there is no shock interaction, many discrete frequencies are present. A broadened low frequency component at $[fD/u_\infty]\sim 2 \times 10^{-2}$, and multiple low-amplitude frequencies between $5 \times 10^{-2} \leq[fD/u_\infty]\leq 2 \times 10^{-1}$ are seen. However, for the shock-laden cavity flow case, the dominant frequency is narrowly tuned to $[fD/u_\infty] \sim 2.5 \times 10^{-1}$ with a small drop ($\sim 35\%$) in the amplitude in comparison with the other.  By changing the cavity dimension and altering the shock interaction zone (while sliding the ramp along the top wall) desired frequency is tuned. However, during the alteration, the resonating cavity and shock oscillations in the duct should be coupled. Such a tunable cavity can be used as a versatile passive control device or a fluidic device in various scenarios.



\section{Limitations and future scope}\label{sec:scope}
\sk{In the present study, the findings are only argued from the two-dimensional simulation where the separated flow instabilities are dominated only by the longitudinal modes. However, in reality, the actual flow contains three-dimensional effects even if the cavity is sufficiently wide (large aspect ratio) and bounded due to the lateral instabilities\cite{DiCicca2013,Das2013,Aradag2010,Woo2008}. Similarly, cavities in axisymmetric bodies or bodies of revolution, the helical modes\cite{Abdelmwgoud2020} in the cavity shear layer change the overall dynamics. The underlying physical events might be similar as described in $\S$\ref{sec:generic_diff}. However, the generic outcomes will relatively change, especially in terms of the dominant spectra and the corresponding power spectral density. Hence, a thorough experimental campaign or a rigorous high-fidelity computation is necessary to understand the resulting cavity dynamics from the shock-shear layer interactions. Similarly, a vital parametric study including the influence of impinging ramp-generated shock on the cavity shear layer, outcomes from multiple shock impingement on the cavity shear layer, freestream Mach number variations, cavity aspect ratio effects, implications of cavity depth, heat addition or release consequences, and repercussions of back-pressure forcing is needed to characterize the resulting flow field completely. Owing to the present paper's limited scope, they are not discussed and left as future research scope.}

\section{Conclusions}\label{sec:conc}
A two-dimensional confined supersonic cavity is numerically investigated. A commercial flow solver is used, and the flow field is resolved using DES-based simulation for the experimental conditions of Panigrahi \etal \cite{Panigrahi2019}. The influence of impinging shock strength on the cavity-free shear layer is particularly studied. Impinging shocks of five different strengths are generated using a small ramp on the top-wall: $[p_2/p_1]=1.0,1.2,1.5,1.7$ and $2.0$. The resulting shock dynamics inside the cavity and the duct are monitored. Following are the major conclusions of the present study:

\begin{enumerate}
    \item {From the unsteady statistics of static-pressure distribution inside the cavity wall, strong harmonics at $[p_2/p_1] = 1.5$ and $1.7$ are observed. Rossiter modes corresponding to $n=1$ suddenly disappear once the shock-shear layer interaction begins and strong harmonics are seen ($n=2$ and $4$). At $[p_2/p_1]=2.0$, the harmonics also disappear, and strong shock interactions from the Mach reflection drive the flow. A maximum rise of 25\% in the time-averaged wall-static pressure is observed between the no-shock and shock interaction cases.}
    \item {Through the $x-t$ plots of static pressure distribution along the cavity-wall and on the freestream probing rakes, the presence of periodic patterns in particular for $[p_2/p_1]=1.5$ case is identified. The periodic patterns are attributed to the disappearance of $n=1$ mode and the presence of strong harmonics. The reason is the synchronous forcing of vortical structures by the impinging shock and the resulting transverse wave-front emission being in resonance with the reflecting waves inside the cavity.}
    \item{The resonance is also found to be coupled with the shock oscillations inside the duct, in particular for the $[p_2/p_1]=1.5$ case. The pressure pulses emitted from the cavity along the leading edge and the fluid mass ejected along the trailing edge primarily drive the cavity coupled shock motion inside the duct. The behavior is only seen for the $[p_2/p_1]=1.5$ case owing to the interaction of wave-fronts almost on the mid-section of the cavity floor. For other cases, due to the weak shock reflections or strong Mach reflections, the impinging shock intercepts the shear layer well upstream. The resulting asymmetric wave-front interaction with the cavity-floor breaks the resonance.}
    \item {From the modal analysis, it is also shown that the dominant spatial mode remains almost the same until $[p_2/p_1]=1.5$. If the freestream contains shock strength larger than the specified, then the dominant spatial mode between the no-shock and shock-shear layer interaction cases are observed to be different. The finding has direct implications in quantifying experimental uncertainties or a maximum permissible shock strength in confined supersonic cavity flows.}
    \item {\sk{The coupling of cavity resonance and shock-oscillation in the duct is exploited further to device a flow control technique. A confined supersonic wall-jet with a cavity upstream is used to demonstrate the passive flow control capabilities. Shock-shear layer interaction in the cavity reduces the jet-column instability and fluctuations by 58\% in the entrained flow by producing periodic large-scale structures along the shear layer. Static pressure spectral analysis along a point in the shear layer between the no-shock and shock-laden cavity flow shows tuned frequency (presence of distinct narrow-band frequency instead of multiple discrete frequencies).}}
\end{enumerate}

\section*{Supplementary material}
See supplementary material for a video showing the time-varying spatial contours of $[p/p_\infty]$ and $[\rho/\rho_\infty]$ for all the cases of shock-shear layer interactions. The respective file is available under the name \href{https://youtu.be/5aa3lQNFo8A}{`video.mp4'}.

\section*{Acknowledgment}
The author would like to thank the host laboratory for permitting him to finish compiling the independent research at the time of corona lock-down, especially to Prof. Jacob Cohen for his continuous encouragement and support. The author thanks the Technion Post-Doctoral fund given in parts with the Fine Trust. The author expresses his deep gratitude to his wife, Dr. Hemaprabha Elangovan, and his colleagues Dr. Purushothaman Nandagopalan and Dr. Senthil Kumar Raman for their rigorous internal reviews and insights provided during the computations and analysis. \sk{Finally, the author thanks all the three reviewers for their appreciation and constructive comments, which helped in reshaping the manuscript to a significant extent.}

\section*{Data availability statement}
The data that support the findings of this study are available from the corresponding author upon reasonable request.

\section*{References}
\bibliography{References}
\
\onecolumngrid
\PRLsep
\end{document}